\documentclass{aa}
\usepackage{natbib}
\usepackage{graphicx}
\usepackage{txfonts}
\begin{document}

   \title{The distribution of stars around the Milky Way's
   central black hole:\\ II. Diffuse light from sub-giants and dwarfs}
  \titlerunning{Distribution of stars around Sgr\,A*}
 \authorrunning{Sch{\"o}del et al.}


   \author{R. Sch\"odel
          \inst{1}
          \and
           E. Gallego-Cano
          \inst{1}
          \and
          H. Dong
          \inst{1}
          \and
          F. Nogueras-Lara
          \inst{1}
          \and
          A. T. Gallego-Calvente
          \inst{1}
          \and
          P. Amaro-Seoane
          \inst{2}
          \and
          H. Baumgardt
          \inst{3}
          }

   \institute{
    Instituto de Astrof\'isica de Andaluc\'ia (CSIC),
     Glorieta de la Astronom\'ia s/n, 18008 Granada, Spain
              \email{rainer@iaa.es}
         \and
       Institut de Ci{\`e}ncies de l'Espai (CSIC-IEEC) at Campus UAB, Carrer de Can Magrans s/n 08193 Barcelona, Spain\\
      Institute of Applied Mathematics, Academy of Mathematics and Systems Science, Chinese Academy of Sciences, Beijing 100190, China\\
   Kavli Institute for Astronomy and Astrophysics, Beijing 100871, China\\
 Zentrum f{\"u}r Astronomie und Astrophysik, TU Berlin, Hardenbergstra{\ss}e 36, 10623 Berlin, Germany
         \and
      School of Mathematics and Physics, University of Queensland
      St. Lucia, QLD 4068,Australia
             }

   \date{Received; accepted }


  \abstract
  {This is the second of three papers that search for the predicted
    stellar cusp  around the Milky Way's central black hole,
    Sagittarius\,A*, with new  data and methods.}
   {We aim to infer the distribution of the faintest
     stellar population currently accessible through observations
     around Sagittarius\,A*.}
   {We used adaptive optics assisted high angular resolution images
     obtained with the NACO instrument at the ESO VLT. Through
     optimised PSF fitting we removed the light from all detected stars
     above a given magnitude limit. Subsequently we analysed the
     remaining, diffuse light density. Systematic  uncertainties were constrained by the
   use of data from different observing epochs and obtained with different filters.
   We show that it is necessary to correct for the diffuse emission
   from the mini-spiral, which would otherwise lead to a systematically
   biased light density profile. We used a Paschen
   $\alpha$ map obtained with the Hubble Space Telescope for this
   purpose.}
 {The azimuthally averaged diffuse surface light density profile within a projected distance of
   $R\lesssim0.5$\,pc from Sagittarius\,A* can be described
   consistently by a single power law with an exponent of
   $\Gamma=0.26\pm0.02_{stat}\pm0.05_{sys}$, similar to what
   has been found for the surface number density of faint stars in Paper\,I.}
 {The analysed diffuse light arises from sub-giant and main-sequence
   stars with $K_{S}\approx19-22$ with masses of
     $0.8-1.5$\,M$_{\odot}$. These stars can be old enough to be
   dynamically relaxed.  The observed power-law profile and its slope
   are consistent with the existence of a relaxed stellar cusp around
   the Milky Way's central black hole. We find that a Nuker law
   provides an adequate description of the nuclear cluster's intrinsic
   shape (assuming spherical symmetry). The 3D power-law slope
     near Sgr\,A* is $\gamma=1.13\pm0.03_{model}\pm0.05_{sys}$. The
     stellar density decreases more steeply beyond a break radius of
     about 3\,pc, which corresponds roughly to the radius of influence
     of the massive black hole.  At a distance of 0.01\,pc from the
     black hole, we estimate a stellar mass density of
     $2.6\pm0.3\times10^{7}$\,M$_{\odot}$\,pc$^{-3}$ and a total
     enclosed stellar mass of $180\pm30$ \,M$_{\odot}$. These
     estimates assume a constant mass-to-light ratio and do not take
     stellar remnants into account. The fact that a flat projected
     surface density is observed
     for old giants at projected distances
   $R\lesssim0.3$\,pc implies that some mechanism may have altered their
   appearance or distribution. }


   \maketitle
%

\section{Introduction}

The existence of power-law stellar density cusps in dynamically
relaxed clusters around massive black holes (BHs) is a fundamental
prediction of theoretical stellar dynamics. The problem of a
stationary stellar density profile around a massive, star-accreting BH
was first analysed by \citet{Peebles:1972fk}, followed by
\citet{Frank:1976uq}, \citet{Lightman:1977ly}, and
\citet{Bahcall:1976vn}. Eight years before \citet{Peebles:1972fk},
\citet{Gurevich:1964vn} had obtained an analogous solution for the
distribution of electrons in the vicinity of a positively charged
Coulomb centre. Since then, many authors have worked on this problem
with a broad variety of methods and have come to similar conclusions
\citep[see, e.g.][and references
therein]{Amaro-Seoane:2004kx,Alexander:2005fk,Merritt:2006ys}.

The in principle best-suited environment where
we can test the presence of such a cusp is the nuclear star cluster
around the massive black hole at the centre of the Milky Way
\citep[e.g.][]{Genzel:2010fk,Schodel:2014bn}.  Unfortunately, the
observations have been limited to the red clump (RC) stars and
brighter giants so far. The density profile of these stars appears to
suggest the absence of a stellar cusp
\citep{Buchholz:2009fk,Do:2009tg,Bartko:2010fk}. However, these stars
only represent a small fraction of the old stars in the nuclear
cluster. It has been proposed that stellar collisions removed their
envelopes in the innermost, densest regions of the cusp, which would
render them invisible \citep[see, e.g.
][]{Alexander:1999fk,Dale:2009ca}, but this cannot fully explain the
observations. Stars that formed less than a few Gyr ago would not be
dynamically relaxed and could thus display a core structure
\citep{Aharon:2015uq}, but the star counts are dominated by RC stars,
which are typically older than a few Gyr. Also, the star formation
history of the central parsec shows that at around 80\% of the stars
formed more than 5\,Gyr ago \citep{Pfuhl:2011uq}.  Another
possibility, that has been recently put forward, is that they
interacted in the past with (a) fragmenting gaseous disc(s), which is
an efficient way to get rid of their envelopes
\citep{Amaro-Seoane:2014fk}. The results of \citet{Kieffer:2016fk}
partially reproduce the findings of \citet{Amaro-Seoane:2014fk}, but
they focused on more compact stars, typically red clump stars, and
find that more hits are required to strip off the envelope of the
star, as stated in the work of \citet{Amaro-Seoane:2014fk}.

The fact that we are dealing with a fundamental problem of stellar
dynamics, the ambiguity of the observational data and their
interpretation, as well as the implications of stellar cusps for the
frequency of Extreme-Mass Ratio Inspirals (EMRIs, see
\citealt{Amaro-Seoane:2007ve} and the review
\citealt{Amaro-Seoane:2012ty} and references therein), and thus on the
detection rate of sources of gravitational radiation
\citep{Hopman:2005dn}, have urged us to revisit this topic.  In
particular, the L3 mission of the European Space Agency has been
approved to be devoted to low frequency gravitational wave astronomy,
with EMRIs being an important class of potential sources.  The mission
implementing this science will follow the Laser Interferometer Space
Antenna (LISA) mission concept
\citep{Amaro-Seoane:2012ys,Amaro-Seoane:2013zr} or a similar one, like
the Chinese Taiji concept \citep{Gong:2015zr}..

This is the second one of a series of papers addressing the
distribution of stars around Sagittarius\,A* (Sgr\,A*). They are
closely related and use the same data, but focus on different methods
and stellar populations. In this work we use the diffuse light
density, while in our first paper (Gallego-Cano et al.\
arXiv:1701.03816, from now on referred to as Paper I), we analyse the
star counts from the brighter, resolved stellar population. We also
refer the interested reader to the more detailed introduction of
Paper\,I for more details about the history and the current state
of the investigation of the stellar cusp at the centre of the Milky
Way.

Our primary goal is to find the predicted stellar cusp of the nuclear
stellar cluster (NSC) around Sgr\,A*. To reach this aim, we push the
boundaries of observational evidence by reaching towards fainter
magnitudes and thus accessing a more representative sample of stars in
the nuclear cluster. In Paper I, we show how we use stacking and
improved analysis methods to provide acceptably complete star counts
for stars about one magitude fainter than what has been done up to
now. These stars, of observed magnitudes $K_{s}\approx18$ at the
distance and extinction of the Galactic Centre (GC), could be
  sub-giant stars, with masses of $1-2\,M_{\odot}$, and potentially be
  old enough to be dynamically relaxed. Indeed, their distribution
inside of a projected distance of $R\lesssim1.0$\,pc can be
approximated well by a single power-law with a slope of
$\Gamma=0.47\pm0.07$. This finding is consistent with the existence of
a stellar cusp of old stars around Sgr\,A*, as we discuss in
Paper\,I. Here, we focus on the diffuse stellar light density around
Sgr\,A*, which provides us with information on even fainter stars.

\section{Data reduction and analysis \label{sec:data}}

\subsection{Basic reduction}

\begin{figure*}[!htb]
\includegraphics[width=\textwidth]{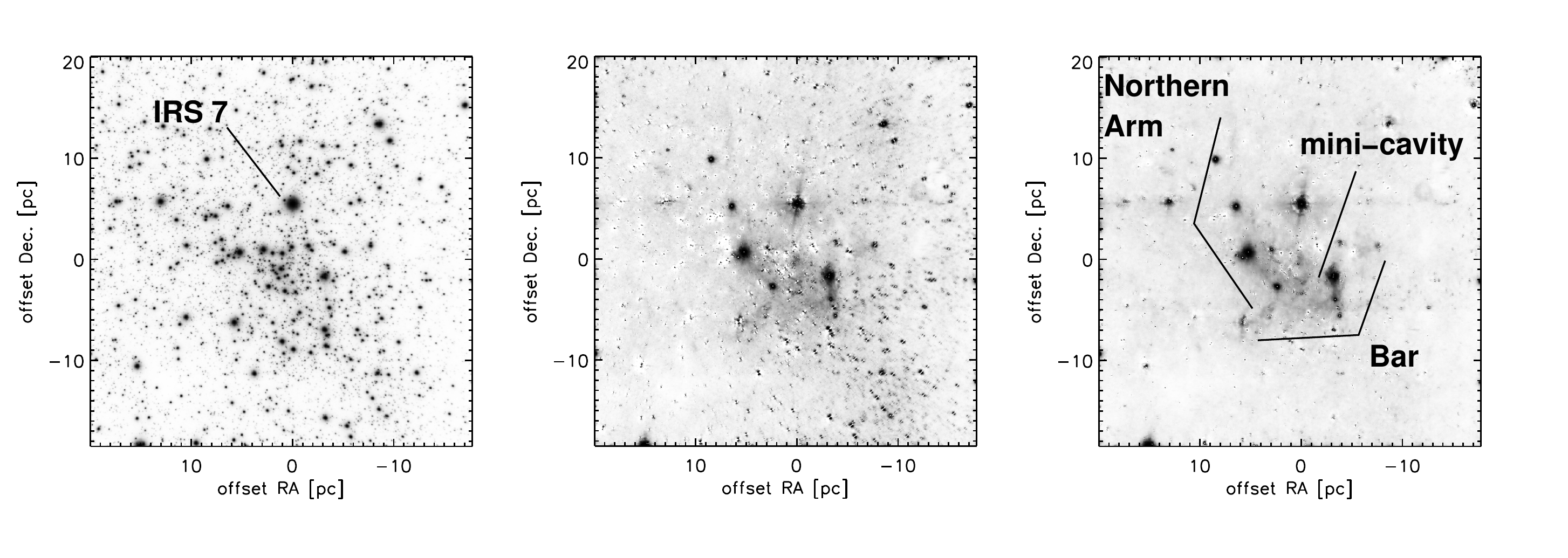}
\caption{\label{Fig:subtraction} Source subtraction. Left: Deep
  $K_{S}$-band mosaic (see Paper\,I). Middle: Deep $K_{S}$-band mosaic
  with all detected stars subtracted, using a single, constant PSF,
  for the entire field. Right: Deep $K_{S}$-band mosaic with all
  detected stars subtracted, using a locally extracted PSF kernel
  merged with a constant halo, that is estimated from IRS\,7. Typical
  features of the mini-spiral of  gas are
  indicated. Logarithmic colour scale in all images, with the same
  scale used in the middle and right panels. North is up and east is
  to the left.}
\end{figure*}

We use the same $H$ and $K_{S}$-band data obtained with the S27 camera
of NACO/VLT that are used in Paper\,I and, additionally, $K_{S}$-band
NACO/VLT S13 camera data from 4 May/12 June/13 August 2011, 4 May/9
August/12 September 2012, and 29 March/14 May 2013. We follow the same
data reduction steps. The S13 images were stacked to provide a deep
image, as done with the S27 images in Paper\,I. In addition, we use
the calibrated HST/NICMOS\,3 image of the emission from  gas at
$1.87\,\mu$m, that was presented by \citet{Dong:2011ff}. We also make
use of NACO/VLT S27 Brackett-$\gamma$ (Br$\gamma$) narrow band (NB)
observations, obtained on 5 August 2009, with a detector integration
time ($DIT$) of 15\,s, 3 averaged readouts per exposure ($NDIT=3$),
and 45 dither positions ($N=45$). Data reduction was standard, as
described in Paper\,I, including rebinning to a finer pixel scale by a
factor of 2. Finally, we use the intermediate-band (IB) filter
imaging data at $2.27\,\mu$m described in Table\,1 of
\citet{Buchholz:2009fk}.

\subsection{Source subtraction}

Subtraction of detected stars is a critical step when estimating the
diffuse light. A particular challenge in AO observations is the
presence of the large seeing halo (FWHM on the order $1"$) around the
near-diffraction limited core of the PSFs. The dynamic range of the
detected stars comprises $>$10 magnitudes, from the brightest star,
GCIRS\,7 with $K_{S}\approx7$ to the faintest detectable stars with
$K{s}\approx19$ (see Paper I).  Many of the brightest stars
($K_{S}=9-11$) are young, massive stars concentrated in the IRS\,16,
IRS\,1, IRS\,33, or IRS\,13 complexes in the central 0.5\,pc
\citep[e.g.][]{Genzel:2003it,Lu:2005fk,Lu:2009bl,Paumard:2006xd}. They
must be carefully subtracted to avoid a bias in the surface light
density.  In addition, the PSF changes across the field due to
anisoplanatic effects, and the variable source density and extinction
mean that the faint wings of the PSFs cannot be estimated with similar
signal-to-noise in all parts of the field because there is not a
homogeneous density of bright, isolated stars.

As explained in Paper I, we extracted the PSFs on overlapping
sub-fields, smaller than the isoplanatic angle. In each of these
sub-fields we used about ten isolated stars -- the brightest ones
possible -- to estimate the PSF core. We then fitted the PSF halo
determined from the brightest star in the field, GCIRS\,7, to the
cores. \citet{Schodel:2010fk} have shown for NACO AO GC data that the
variation of the PSF halo is rather negligible, which means that with
the chosen approach we can reach a photometric accuracy of a few
percent across the entire field. Subsequently, point sources are
detected and subtracted. Since the detection of occasional spurious
sources is no source of concern for this work, we chose a more
aggressive approach than in Paper\,I, setting the {\it StarFinder}
parameters $min\_correlation = 0.70$ and $deblend=1$ for all images
(except if stated explicitly otherwise).  We note that even with these settings the
detection completeness falls below 50\% for sources fainter than about
$K_{S}=18.5$ in the centralmost arcseconds. We note that we only
  perform point-source fitting and subtraction, but do not model the
  diffuse background with {\it StarFinder}, that is, the keywords
  $BACK\_BOX$ and $ESTIMATE\_BG$ are set to zero.

In order to have an extinction map that covers even the large area of
the wide field observations from May 2011, we created an extinction
map from HAWK-I $H$ and $K_{S}$ speckle holography-reduced FASTPHOT
observations of the central square arcminutes (Nogueras-Lara et al.,
in prep.).
We used the extinction law of \citet{Schodel:2010fk}
($A_{\lambda}\propto\lambda^{-2.2}$), assumed a constant intrinsic
colour of $(H-K_{S})_{0}=0.1$ for all the stars, and used the mean of
the 20 nearest stars for each pixel. This results in an extinction map
with a variable angular resolution of roughly $2"$. The results
presented in this paper are not sensitive in any significant way on
the variation in these assumptions within their uncertainties. In
particular, changing the exponent of the extinction law to other
plausible values \citep[e.g. $2.0$, see][]{Nishiyama:2009oj} will
have an impact on any of the parameters of interest that is a factor
of a few smaller than other sources of uncertainties that will be
discussed here.

\begin{figure*}[!htb]
\includegraphics[width=\textwidth]{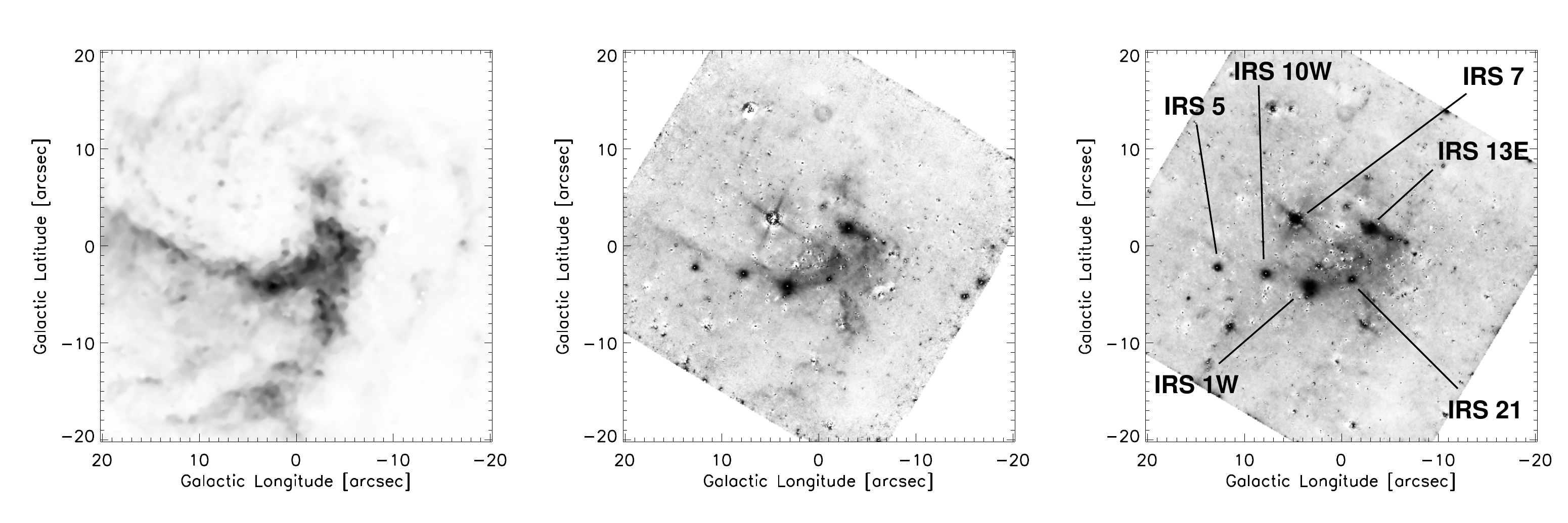}
\caption{\label{Fig:gas} Left: HST NIC3 point-source-subtracted
  Pa\,$\alpha$ image of the GC. Middle: VLT NACO point-source-subtracted
  Br-$\gamma$ image. Right: VLT NACO point-source-subtracted
  $K_{S}$  image. Some prominent point-like emission sources (see
  text) are labelled. }
\end{figure*}

\begin{figure*}[!htb]
\includegraphics[width=\textwidth]{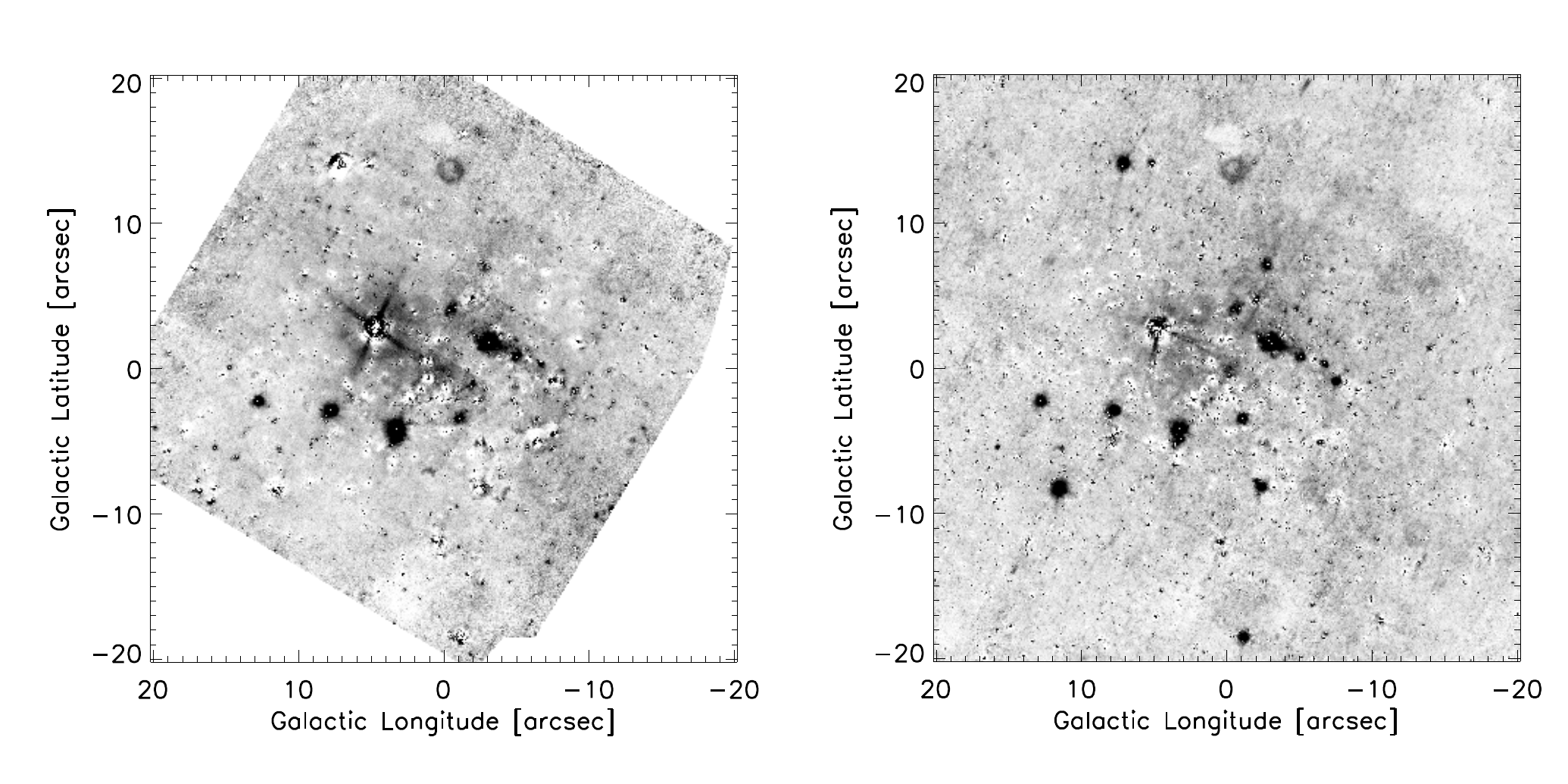}
\caption{\label{Fig:minuspa} Left: Point-source-subtracted Br-$\gamma$
  image minus scaled Pa\,$\alpha$
  image. Right: Point-source-subtracted $K_{S}$ wide field image minus
  scaled Pa\,$\alpha$ image.}
\end{figure*}

We demonstrate the result of this strategy in
Fig.\,\ref{Fig:subtraction}. There, we show the mosaic of the deep
$K_{S}$ image (see Paper \,I), and the same field after subtraction
with a single, constant PSF, and after subtraction with a variable
PSF, composed of a local core plus a global halo. As can be seen,
using a single, constant PSF leads to variable artefacts associated
with the stellar sources across the field \citep[see
also][]{Schodel:2010hc}. Also, when the wings of the PSF are not
determined with high signal-to-noise, then the diffuse emission is
dominated by flux from the seeing halos around bright stars.

With the variable core plus halo PSF (determined from the brightest
star IRS\,7, right panel), the residuals around bright stars are
strongly suppressed and any remaining residuals are largely constant
across the field, as can be seen in the right panel of
Fig.\,\ref{Fig:subtraction}. These remaining residuals are typical for
PSF subtraction with an empirical PSF when the PSF is not fully
constant across the field: Since several stars have to be used to
derive a median PSF, their slightly different PSFs will result in a
slightly too broad median PSF. This leads to the typical and
inevitable artefacts in the form of core-excesses with surrounding
negativities that can be seen around bright stars. Nevertheless, as
can be seen, the residuals around the bright stars have been strongly
suppressed with our method. The only exception is GCIRS\,7, which is
extremely bright (a few magnitudes brighter than any other source in
the field). The filamentary structure of the
so-called mini-spiral \citep[see][and references
therein]{Genzel:2010fk} becomes apparent, with features such as the
northern arm, the bar, or the mini-cavity clearly visible. We provide
further detailed tests of our methodology in Appendix\,\ref{app:photo}.

\subsection{Subtraction of mini-spiral emission \label{sec:gas}}

As we can see in the right panels of Figs.\,\ref{Fig:subtraction} and
\ref{Fig:gas}, diffuse emission from the so-called
mini-spiral \citep[see., e.g.][]{Genzel:2010fk} contributes
significantly to the diffuse emission within about 0.5\,pc
($\sim$$12"$
for a GC distance of 8\,kpc) of Sgr\,A*, even in broad band images. We
therefore have to correct for it before we will be able to estimate
the diffuse emission arising from unresolved \emph{stellar} sources. At
the wavelengths considered, the emission can arise from hydrogen
and helium lines (e.g. HI at $2.17$,
$1.64$
or $1.74\,\mu$m,
HeI at $1.70$,
$2.06$,
or $2.11\,\mu$m), but some contribution from hot and warm dust is also
plausible.
In Fig.\,\ref{Fig:gas} we show the mini-spiral as seen in the
Paschen\,$\alpha$
line with NIC3/HST and in the Brackett$\gamma$
line as well as in $K_{S}$ with NACO/VLT, respectively. The Pa\,$\alpha$
image is from the survey by \citet{Wang:2010fk} and
\citet{Dong:2011ff}.

Since we will use the HST image as a reference for gas emission, we
aligned all our images via a first order polynomial transform with the
HST image. The positions of detected stars were used to calculate the
transformation parameters with IDL POLYWARP and the images were then
aligned using IDL POLY\_2D. The pixel scale of the resulting images is
set to the one of the HST image ($0.101"$ per pixel).

As can be seen in Fig\,\ref{Fig:gas}, the Pa\,$\alpha$ image traces
the gas emission very clearly \cite[with the exception of a few
Pa\,$\alpha$ excess sources, see][]{Dong:2012fk} and the $K_{S}$ and
Br\,$\gamma$ images of the diffuse emission trace the same structures
of the mini-spiral. Some differences are given by residuals around
bright stars, by some residual emission associated with the brightest
star, GCIRS\,7, by hot dust emission around the probable bow-shock
sources IRS\,21, IRS\,10W, IRS\,5, and IRS\,1W, and by enhanced
emission in and around the IRS\,13E complex, probably from a higher
gas temperature. We mark some of these sources and areas in
Fig.\,\ref{Fig:gas} and will mask them when deriving scaling factors
for gas subtraction and when computing the brightness of diffuse
stellar light in the following sections.

Figure\,\ref{Fig:minuspa} shows the point source-subtracted
Br\,$\gamma$ and $K_{S}$ images after subtraction of the
scaled Pa\,$\alpha$ image. The scale factor was assumed constant and
estimated by eye. All images were corrected for differential
extinction. We also determined the scaling factor in a numerical way
fitting the azimuthally averaged surface brightness distribution
  with a least $\chi^{2}$ fit with a linear combination of the
  scaled azimuthally averaged surface brightness distribution of the
  Pa\,$\alpha$ emission plus a simple power-law diffuse light density
  distribution centred on Sgr\,A*.
\begin{equation}
\label{equ:powerlaw}
\Sigma(R) = \Sigma_{0} * (R/R_{0})^{-\Gamma} + \beta Pa\,\alpha,
\end{equation}
where $\Sigma_{0}$ is the surface flux density at a the projected
distance $R_{0}=0.5$\,pc, $\Gamma$ is the power-law index, Pa\,$\alpha$ is the
Paschen\,$\alpha$ SB, and $\beta$ the scaling factor for the latter.
There are three free parameters, $\Sigma_{0}$, $\Gamma$, and $\beta$ .
We limited the estimation of $\beta$ to the region $R\leq0.5$\,pc,
where the gas emission is strongest. Varying this value up to
$R=1.5$\,pc does not have any significant effects on the SB profiles,
but some negativities can then appear after minispiral subtraction in the images
because the fit is dominated by regions at large $R$ with low
 gas SB, where the excitation conditions of the gas may also
be different (greater distance from the hot stars near Sgr\,A*). The
resulting best-fit factor was close to our by-eye estimate.

We considered fitting the azimuthally averaged surface brightnesses as
marginally more reliable than directly fitting the images because the
azimuthal average will suppress noise from the data acquisition and
reduction process, from the point-source subtraction, and from
potential variations of the gas temperature. Nevertheless, we also tested
direct fitting of the models to the images and obtained the same
results within the formal uncertainties of the fits. As can be seen in
Fig.\,\ref{Fig:minuspa}, most of the emission from the gas and dust in the
mini-spiral can be effectively removed by this simple procedure. From
our by-eye fit we estimated an uncertainty of $10\%$ for the best
scale factor, while its uncertainty from the least $\chi^{2}$ fits is
$<5\%$. This uncertainty has a negligible effect on the parameters we
are interested in, in particular the slope of the power-law surface
density.  For all images and wavelengths used in the following we
applied the numerical procedure to estimate the scaling factor for the
subtraction of the diffuse gas emission.

An alternative way of subtracting the mini-spiral emission may be by
using the intrinsic line ratio of $Br\,\gamma/Pa\,\alpha$. However,
this is not practical in our case because most of our data are
broad-band observations and include additional lines, for example from the
$2.058\,\mu$m He\,I line in the $K_{S}$-band. Also, in the  case of the
$Br\,\gamma$ image, no accurate calibration was possible because no
zero point observations were taken at the time of the
observation and the sky conditions were not photometric.

Finally, NIR emission from the mini-spiral may also arise, at
  least partially, from hot dust, in particular near young, massive
  stars, such as the IRS\,13 region or the putative bow-shock sources
  IRS\,21, IRS\,1W, etc. \citep[see, e.g.
  ][]{Eckart:2004fk,Fritz:2010tk,Sanchez-Bermudez:2014fk}. This is
  plausible because the morphology of the emission from warm/hot dust
  in the mini-spiral region resembles closely the one observed in line
  emission \citep[compare, e.g. the images of the mini-spiral seen
  through different filters
  in][]{Muzic:2007kx,Wang:2010fk,Genzel:2010fk,Lau:2013fk}. We did
  some experiments in this respect, with point-source subtracted
  $8.6\,\mu$m and $3.8\,\mu$m imaging data \citep{Schodel:2011uq} and
  found dust temperatures on the order of 250-350\,K. This is hotter
  than in the SOFIA observations analysed by \citet{Lau:2013fk} and is
  probably related to us using data of considerably higher angular
  resolution and considerably shorter wavelengths or to the fact that
  the $8.6\,\mu$m image may be dominated by emission from PAHs.

When we correct the measured
  diffuse SB profile in $K_{s}$ with our dust emission map, we get
  roughly similar results than with the HST Pa\,$\alpha$ image. However,
  the quality of the correction is considerably worse because (a) the
  HST data provide much cleaner measurements of the diffuse gas
  emission, (b) the FOV of our $8.6\,\mu$m and $3.8\,\mu$m imaging
  data is smaller than the one of the HST images, (c) point sources
  must first be subtracted from the NIR/MIR images, which introduces
  additional systematic errors, and (d) further systematics are
  introduced by the very challenging determination of the variable sky
  background in the MIR observations (it is impossible to chop into an
  emission-free region inside the GC). The fundamental assumption in
  our work is that the non-stellar diffuse emission can be obtained
  from the HST Paschen\,$\alpha$ image through applying a constant
  scaling factor. As long as this assumption is approximately valid,
  it does not really matter whether we are dealing with line emission
  and/or dust emission. A study of variable line-emission and variable
  gas or dust temperature in the mini-spiral is beyond the scope of this
  paper. As we show below, our method to remove the non-stellar
  diffuse emission appears to work very well and provides consistent
  results across many filters. We therefore believe our method to be solid.

\section{The surface density of faint stars in the GC}

In this section we explore the surface brightness (SB) profile of the
diffuse stellar light in observations taken with different cameras and
filters, as well as at different epochs. We will also perform various
checks on potential sources of systematic bias.

\subsection{Wide field $K_{S}$ band \label{sec:wide}}

First, we examine a wide-field mosaic that was obtained with NACO/VLT S27
in May 2011. In total, $4\times4$ pointings were observed in $K_{S}$, centred
approximately on Sgr\,A*. The images are relatively shallow, with a
total on-target exposure time of only 72\,s per pointing (4 exposures
with $DIT=2s$, $NDIT=9$), but of excellent and homogeneous
quality.

\begin{figure}[!htb]
\includegraphics[width=\columnwidth]{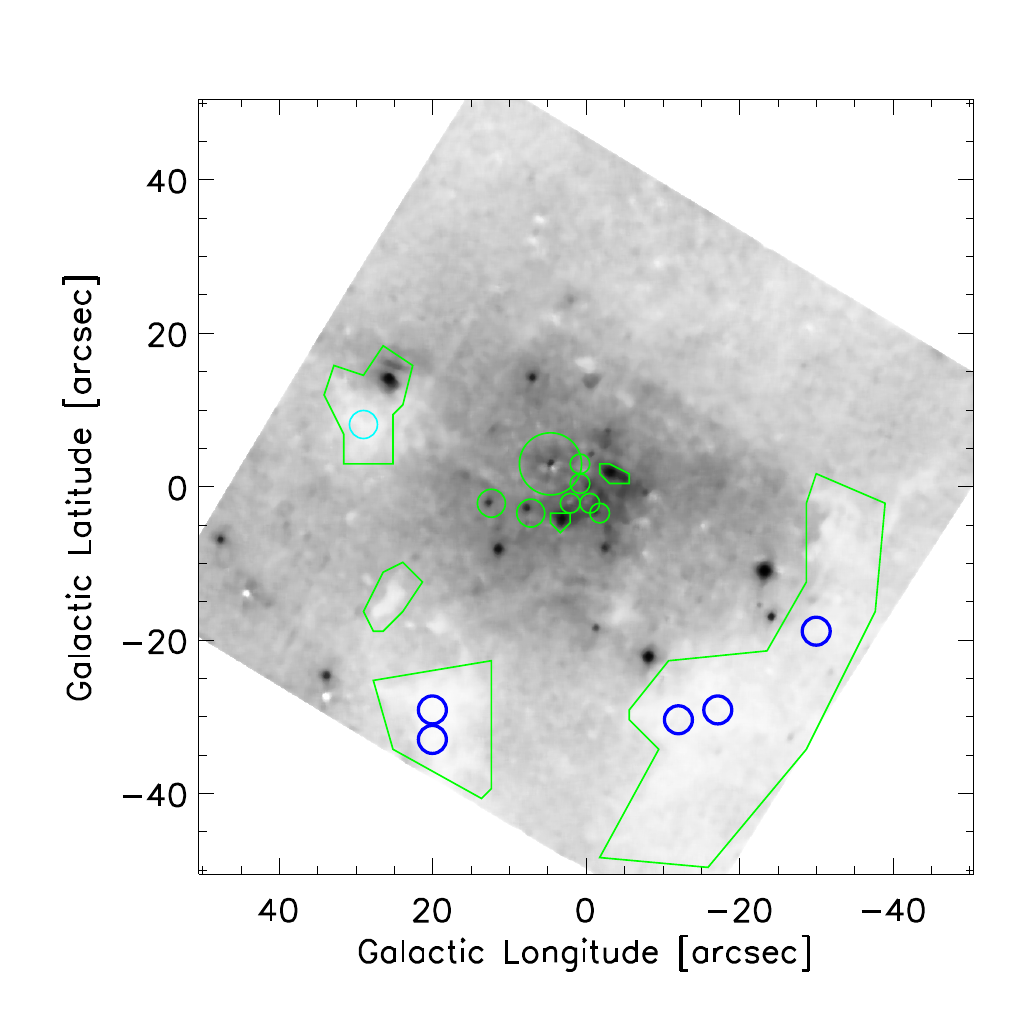}
\caption{\label{Fig:wide} Point-source-subtracted $K_{S}$ wide
  field  image.  Contaminated regions (residuals from bright
  stars, hot dust, and IRS\,13) and dark clouds,
  that are excluded from measuring the surface light density, are
  indicated by green polygons. The blue circles indicate regions that
  were used to estimate the offset of the diffuse flux density.}
\end{figure}

Figure\,\ref{Fig:wide} shows the point-source subtracted
  wide-field image. As mentioned above, we assumed that the diffuse
  light from the stars follows a power law and that a constant scaling
  factor is adequate to remove the emission from the mini-spiral. That
  is, our model is described by Equation\,\ref{equ:powerlaw}. We
  measured the mean diffuse emission in one pixel wide annuli around Sgr\,A*, using
  the IDL ASTROLIB routine ROBUST\_MEAN, rejecting $> 5\,\sigma$
  outliers. The corresponding uncertainties were taken as the
  uncertainties of the means. The same was done for the
  Paschen\,$\alpha$ image. Subsequently, we used a least $\chi^{2}$ fit
  to determine the best parameters to scale the gas emission and
  determine the power law emission for the stars.

\begin{figure}[!htb]
\includegraphics[width=\columnwidth]{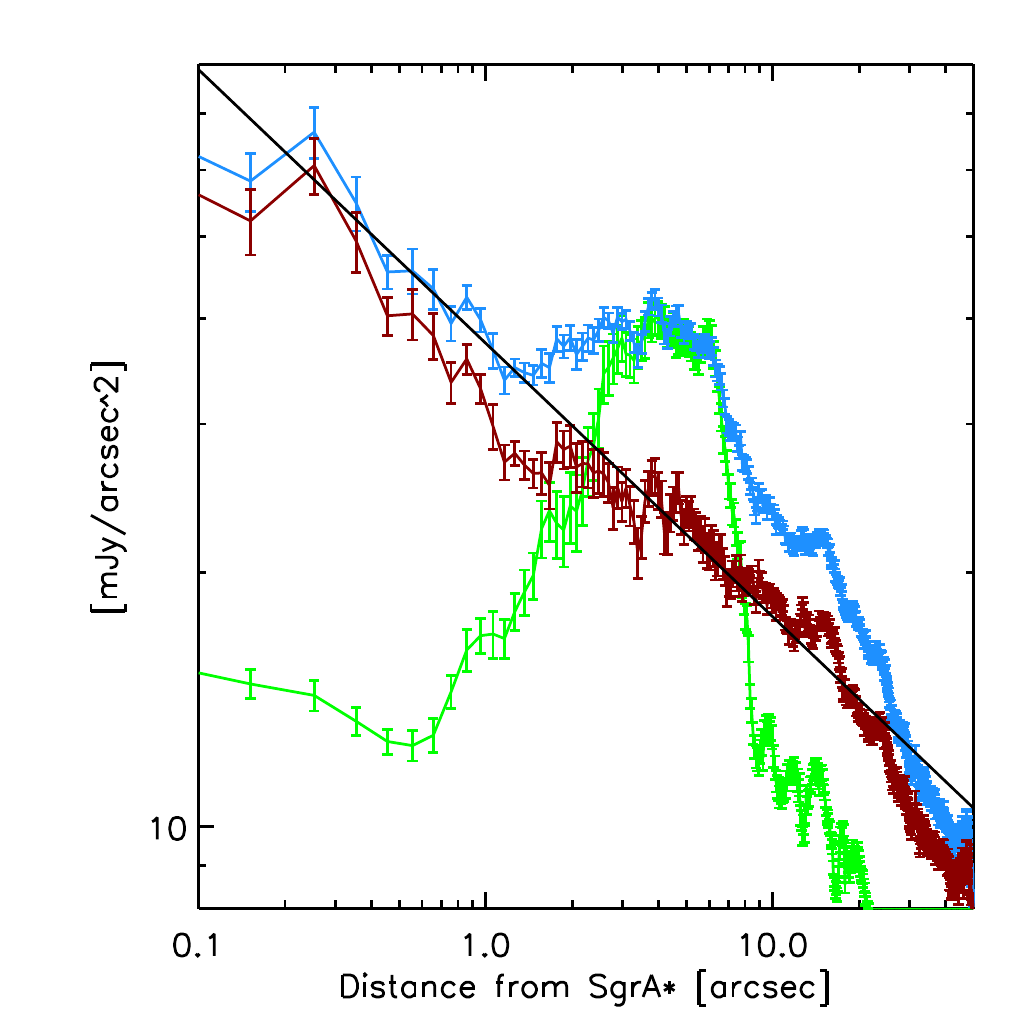}
\caption{\label{Fig:SB_WIDE} Left: Mean diffuse SB profiles in the
  $K_{S}$ wide-field image before (blue) and after (red) subtraction of
the appropriately scaled Pa\,$\alpha$ emission (green; multiplied by
arbitrary factor to optimise the plot). The straight black line is
the best power-law fit to the red data within $R\leq25"$\,pc
(corresponding to $R\lesssim1$\,pc for a GC distance of 8\,kpc).}
\end{figure}

Figure\,\ref{Fig:SB_WIDE} shows the measured surface brightness (SB)
profile for the wide-field $Ks$-image, for the Paschen\,$\alpha$
emission, and the wide-field SB profile after a scaled subtraction of
the latter. The continuous black line is the best-fit power law to the
data at $R\leq1.0$\,pc. It has a reduced $\chi^{2}=14.3$,
  $\Sigma_{0}=16.4\pm0.1$\,mJy\,arcsec$^{-2}$, and
  $\Gamma=0.32\pm0.01$. The relatively high $\chi^{2}$ is mainly due
to systematic deviations of the profile from a power-law at certain
restricted ranges of $R$. We found that these deviations are mainly
related to the difficulties of precise subtraction of bright stars at
small $R$. These systematics are slightly different for each data set
that we present in this work (see, e.g. Fig\,\ref{Fig:SB}), but do
not significantly affect the overall result. The formal
  uncertainties resulting from the fit code have been rescaled to a
  reduced $\chi^{2}=1$ here and for all other fits reported in this
  paper. In appendix\,\ref{sec:app2} we study several potential
sources of systematic errors, such as sky offset, binning, fitting
range, or application of the extinction correction. The sky offset, a
probable systematic effect from inaccurate sky background subtraction
and diffuse foreground (i.e.\ nor originating within the GC), was
estimated on small regions of dark clouds (see Fig.\,\ref{Fig:wide})
and subtracted prior to measuring the SB on the wide field image.

\subsection{Br\,$\gamma$}

\begin{figure}[!htb]
\includegraphics[width=\columnwidth]{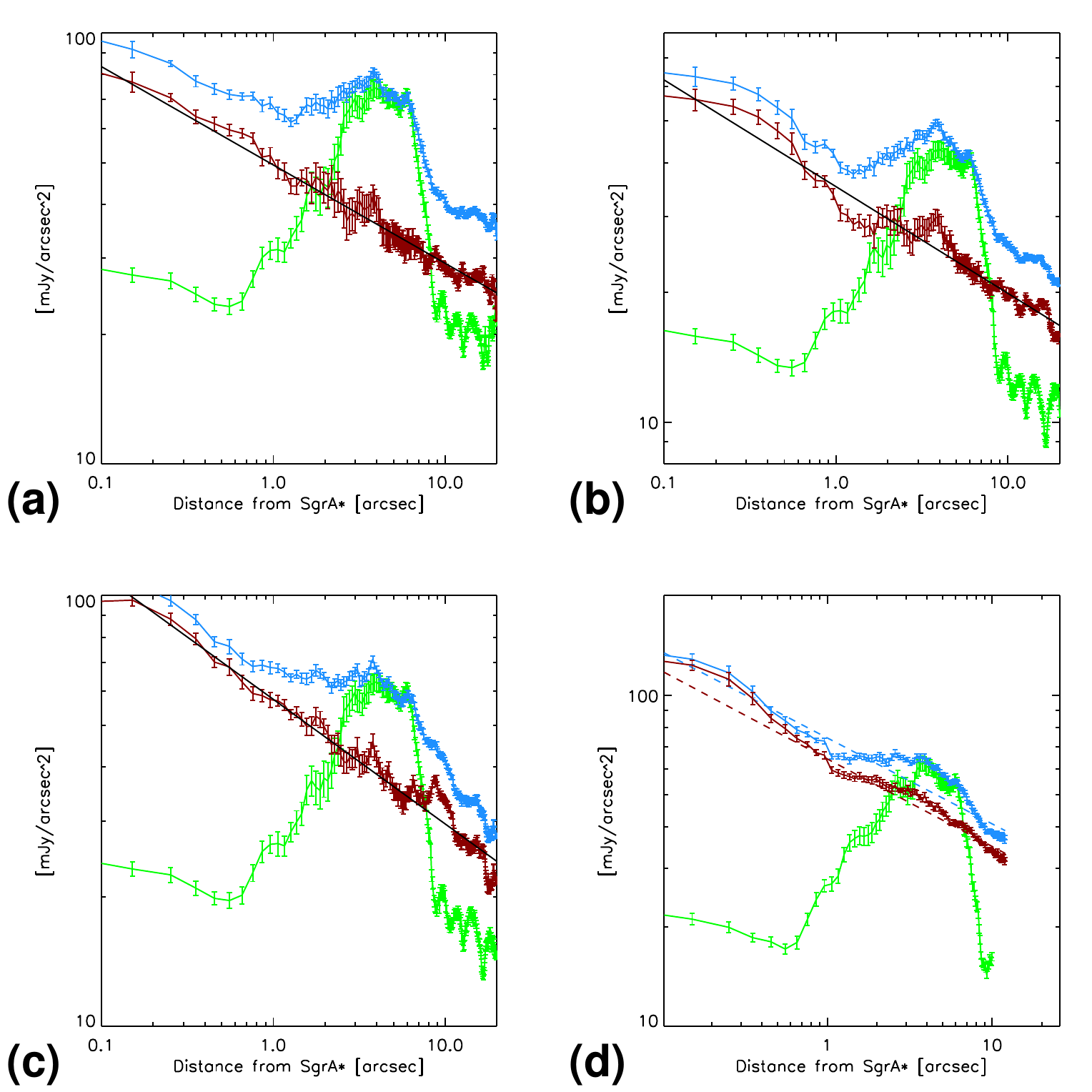}
\caption{\label{Fig:SB} {\bf (a)} Plot of the $Br\gamma$ diffuse SB
  profile before (blue) and after (red) subtraction of the
  appropriately scaled Pa\,$\alpha$ emission (green; multiplied by
  arbitrary factor to optimise the plot). The black line is for the
  best-fit power-law in the range $R\leq25"$\,pc (corresponding to
  $R\lesssim1$\,pc for a GC distance of 8\,kpc). {\bf (b)} As (a), but
  for the deep $K_{s}$ image.  {\bf (c)} As (a), but for the $H$
  image.  {\bf (d)} As (a), but for the $K_{s}$ S13 image.}
\end{figure}

The surface brightness in the Br\,$\gamma$ narrow band filter image is
an interesting test case because here the emission from the ionised
gas will provide a relatively large fraction of the overall diffuse
emission. The resulting raw and ionised-gas-corrected SB profiles are
shown in panel (a) of Fig.\,\ref{Fig:SB}.  A simple power-law
provides a very good fit, with the best-fit power-law exponent of
$\Gamma_{in}=0.23\pm0.01$.

\subsection{Deep $K_{S}$-band image}
Here, we analyse the deep $K_{S}$ broad band image that we use for
measuring the stellar number surface density in Paper I. The SB
profiles are shown in panal (b) of Fig.\,\ref{Fig:SB}.  A simple power-law
provides a very good fit, with the best-fit power-law exponent of
$\Gamma_{in}=0.25\pm0.01$.

\subsection{$H$-band image}

Analysing the diffuse flux in an $H$ band image represents, among
others, a test in a regime, where differential extinction is stronger,
where the sky background behaves in a different way, and where the
ratio of line emission relative to Pa\,$\alpha$ is different. Also, due
to increased anisoplanatic effects, point-source-subtraction removal
is more difficult in $H$ than in $K_{s}$. Hence, the $H$-band can be
very helpful in constraining systematic effects. We had to correct the
$H$-band image for a systematic negative offset of the sky background,
which could be measured on some small dark clouds in the field.  The
SB profiles are shown in panel (c) of Fig.\,\ref{Fig:SB}.  A simple
power-law provides a good fit, with the best-fit power-law exponent of
$\Gamma_{in}=0.29\pm0.01$.

\subsection{Deep $K_{S}$ image with S13 camera}

As a final test, we examine the diffuse light density in a deep,
multi-epoch $K_{S}$-band image obtained with data from the S13 camera
of NACO/VLT. A simple power-law provides a good fit,
with the best-fit power-law exponent of $\Gamma_{in}=0.26\pm0.01$. We tested
again the systematics of subtracting the stars down to different
limiting magnitudes ($K_{s,lim} = 16,18,20$). The power-law index
changes between $0.25\pm0.01$ and $0.29\pm0.01$ and the plot looks similar in all
cases (not shown).  Compared to the NACO S27 $K_{s}$ data there appears
to be an offset of the SB towards brighter values. We could not
identify the source of this offset, but we note that it does not affect
our main conclusions, in particular the existence of a power-law cusp
and its index.

\subsection{IB227 image}

As a final test we used the IB227 image from
\citet{Buchholz:2009fk}. A simple power-law provides a good fit, with
the best-fit power-law exponent of $\Gamma_{in}=0.19\pm0.01$. We do not show
the corresponding fit in Fig.\,\ref{Fig:SB} to not overcrowd the
plot. It is very similar to all the other plots.

\section{Discussion}

\subsection{Mean projected power-law index \label{sec:Gamma}}

\begin{table}
\centering
\caption{Best-fit power law indices for the diffuse stellar light
  inside of $R<0.5$\,pc. All \emph{formal} uncertainties are
  $\leq0.01$ after having been rescaled to a reduced $\chi^{2}=1$ , i.e.\ we are dominated by systematics.}
\label{Tab:Gammas}
\begin{tabular}{ll}
\hline
\hline
Data & $\Gamma_{in}$ \\
\hline
$K_{S}$, wide field  &  0.32\\
Br\,$\gamma$ &  0.23\\
$K_{S}$ deep field &  0.25\\
$H$ &  0.29\\
$K_{S}$ S13 &  0.26\\
IB227  &  0.19\\
 \hline
\end{tabular}
\end{table}

\begin{figure}[!htb]
\includegraphics[width=\columnwidth]{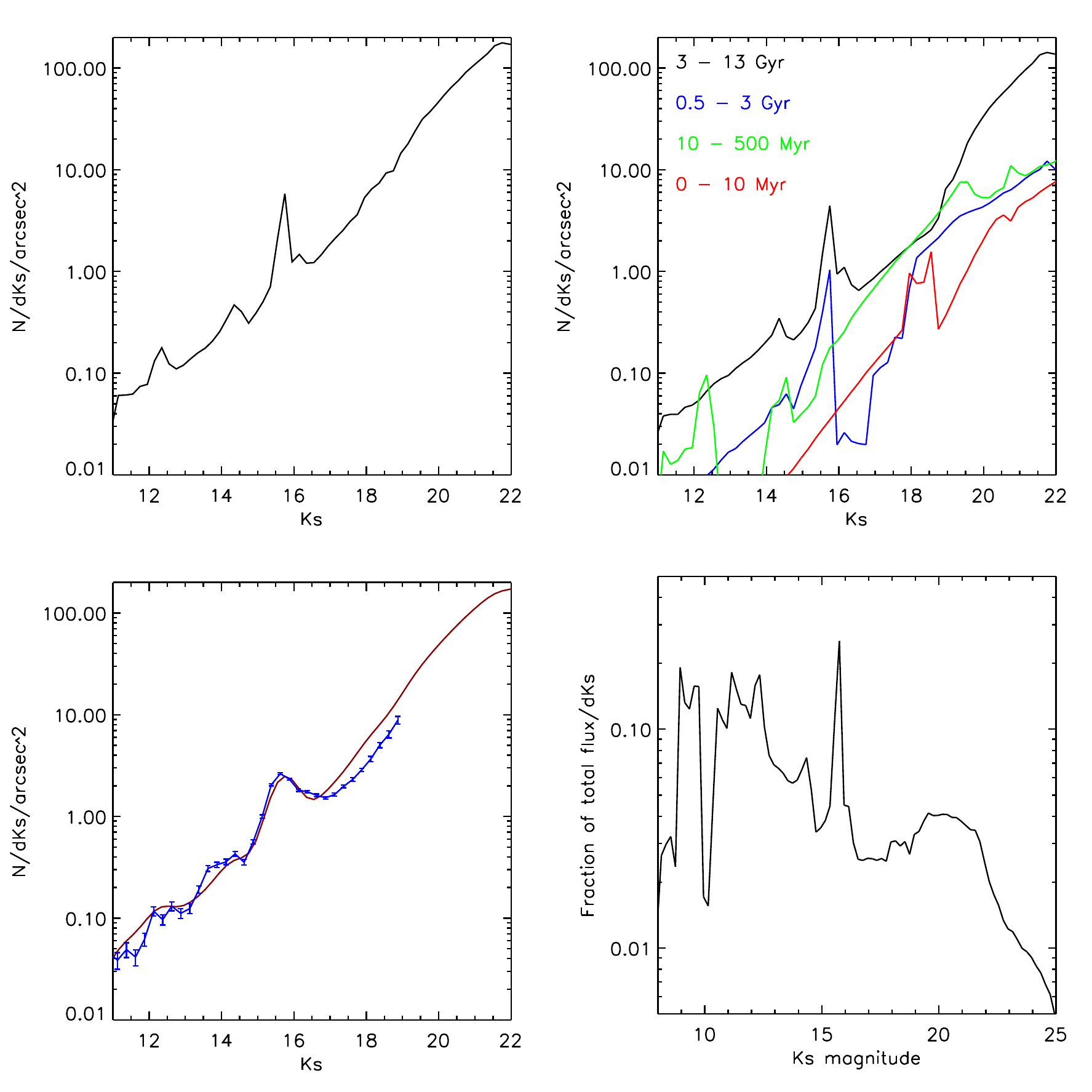}
\caption{\label{Fig:KLF} Estimates of the KLF within $R\leq 1$\,pc of
  Sgr\,A*. Upper left: Model $K_{S}$ luminosity function (KLF) based on
  the star formation history derived by \citet{Pfuhl:2011uq}. Upper
  right: KLFs for stellar populations in certain time windows, using
  the star formation model by
  \citet{Pfuhl:2011uq}.  Lower left: The red line is the model
  KLF smoothed with a Gaussian kernel to roughly take differential extinction into account.  The blue
  line with error bars is the KLF as determined in Paper\,I.  Lower
  right: Fraction of total flux contributed by stars of
  different magnitudes, i.e.\ the KLF multiplied by the flux density
  of stars in a given bin and divided by the total flux. Please note
  the different range of the x-axis in this plot, which is chosen to
  show the decrease of the flux contribution at $K_{s}>22$.}
\end{figure}

As the preceding sections have shown, measuring the diffuse stellar
light around Sgr\,A* is a non-trivial undertaking and subject to
potentially significant systematic effects. In particular, we have
demonstrated that the mini-spiral contributes significantly to the
measured diffuse flux at projected distances $R<0.5$\,pc from Sgr\,A*,
even when broad-band filters are used. If not taken into account,
  this will result in an apparent steep increase of the diffuse flux
at $R\lesssim0.4$\,pc and then an almost flat SB profile in the
innermost $\sim$$0.2$\,pc.
The exact systematic effect due to the mini-spiral will depend, of
course, on the filter used. It is very strong in
Br$\gamma$ and weaker in $K_{S}$ and $H$ (see Fig.\,\ref{Fig:SB}).

The subtraction of the flux of both the bright stars and the
gas and dust is prone to systematic errors. Fortunately, these errors will
change with the observing conditions, for example seeing and adaptive optics
correction, camera used, or observing wavelength. For this reason, we
have used several completely independent data sets that were obtained
at different times and with significantly different setups: Deep and
shallow images, broad and narrow band observations, shorter and longer
wavelength filters. It is satisfying to see that the resulting SB
profile is consistent among all the data sets.

All our different measurements of the projected stellar surface
brightness can be fit well by the simple model of a single power-law
at $R\lesssim1$\,pc. The corresponding power-law indices are
consistent with each other and also with the power-law index inferred
for the stellar number density of faint stars in this region, as
determined in Paper\,I.  As studied and explained in detail in
appendix\,\ref{sec:app2}, the systematic error is dominated by effects
of potential additive sky offsets and fitting range. The atmospheric
contribution of the former is, however, variable in nature between
sets of different observations and is therefore absorbed into the
statistical error from the mean of the different values for $\Gamma$
observed. The latter is mainly caused by a systematic steepening of
the slope with increasing $R$ and contributes an estimated $0.05$ to
the uncertainty budget of $\Gamma$. As concerns the contribution of a
potential source of diffuse emission from a stellar foreground
population, for example in the nuclear disc, we do not take it into account
here. We note, however, that its contribution would always be an
additive offset. If taken into account, this would systematically
steepen the observed $\Gamma$.

Table\,\ref{Tab:Gammas} lists the resulting best-fit power-law indices
for the projected diffuse light in the inner $1.0$\,pc. From
these measurements to independent data sets we obtain a mean estimate of
$\Gamma=0.26\pm0.02_{stat}\pm0.05_{sys}$. This value is smaller than,
but agrees within
its uncertainties, with what we observe for the number density of the
stars in the range $17.5\lesssim K_{S} \lesssim 18.5$ that we present
in Paper\,I. We conclude that the projected surface density
distribution of stars around Sgr\,A* can be described well by a single
power law with the same exponent for different stellar
populations. The faint stars do not show a flat, core-like
distribution as has been observed for the bright ($K_{S}\lesssim15.5$)
giants in the GC \citep{Buchholz:2009fk,Do:2009tg,Bartko:2010fk}. \emph{The faint stellar population around Sgr\,A* clearly displays a
  power-law cusp in the central parsec.} Given our measurements,
assumptions, and analysis, we can exclude a flat projected core around
Sgr\,A* with high confidence. Also, we do not find it necessary to use
any broken power-law for the SB profile at projected radii
$R\leq 1$\,pc, as it was used by previous authors
\citep[e.g.][]{Genzel:2003it,Schodel:2007tw,Do:2009tg}.

\subsection{What kind of stars contribute to the diffuse SB? \label{sec:masses}}

  The stars that contribute dominantly to the diffuse light in our
  point-source subtracted images must be fainter than $K_{S}=18$. To
  obtain a better understanding of which kind of stars contribute
  dominantly to the diffuse SB, we study the $Ks$ luminosity function
  (KLF). First, we use the star formation history for the central
  parsec derived by \citet{Pfuhl:2011uq} to construct a theoretical
  KLF. We used their Eq.\,(3) to compute the masses of nine single
  age stellar populations. The ages were taken to be the middle of the
  intervals $10-13$\,Gyr, $8-10$\,Gyr, $3-8$\,Gyr, $1-3$\,Gyr,
  $0.5-1$\,Gyr, $200-500$\,Myr, $50-200$\,Myr, $10-50$\,Myr, and
  $0-5$\,Myr. We emphasise the illustrative nature of our model, which
  is not constructed to provide a precise fit to our data.

 The model KLFs were calculated assuming Solar metallicities and
Chabrier lognormal initial mass functions \citep[see, {\it
  http://stev.oapd.inaf.it/cgi-bin/cmd\_2.8} and
][]{Chabrier:2001yg,Bressan:2012xy,Chen:2014nr,Chen:2015sf,Tang:2014rm}.

The resulting total KLF and the individual contributions of the
populations of different ages (where we summed over four broad age
ranges) can be seen in the upper panels of Fig.\,\ref{Fig:KLF}. The
lower left panel compares the smoothed (to take into account
differential extinction and measurement uncertainties) model KLF to
the completeness corrected KLF determined by us in Paper\,I. The
agreement is satisfactory. The peaks around $K_{S}\approx15.5$ arise
from Red Clump (RC) stars.  We point out that we have not made any
specific effort to match the model KLF to the measured one, except for
applying a scaling factor. Studies of star formation history or
metallicity are beyond the scope of this paper.

The bottom right panel in Fig.\,\ref{Fig:KLF} shows the fraction of
the total flux contributed by the stars in the different bins of the
model KLF.  As can be seen, stars in the regime $K_{S}=19-22$ do not
differ significantly in their overall weight. We expect these stars,
to dominate our measurements of the diffuse light density.  As can be
seen in the upper right panel, these stars belong predominantly to the
oldest stellar population.  They will be of type G to F,
have masses $\lesssim1.5\,M_{\odot}$ and will live for several Gyrs
\citep[see also Fig.\,16 in][]{Schodel:2007tw}. They can thus be old
enough to be dynamically relaxed and serve as tracers for the
existence of a stellar cusp.

In Paper\,I we discuss and take into account the possible
contamination of the surface number density of $K_{s}\approx18$ stars
by young stars from the most recent, $\sim$5\,Myr-old star formation
event in the central $R<0.5$\,pc \citep[see,
e.g.][]{Genzel:2010fk,Lu:2013fk}. It turns out to be relatively
minor, but the contamination from other young or intermediate-age
populations, with ages $\lesssim3$\,Gyr may be significant. Here we
want to explore whether such contamination could, in principle, also
be present in the diffuse light. While we cannot completely rule out
this possibility, the top right panel of Fig.\,\ref{Fig:KLF} shows
that stars older than a 3\,Gyr will be a factor of a few more frequent
than younger stars at $K_{s}>20$ (see also Fig.\,11 in Paper\,I).

Also, although the stellar number density profiles derived in Paper\,I and
the SB profile measured in this work probe different stellar masses
and ages, the corresponding values of the power-law indices are
approximately consistent with each another. Hence, while it is difficult to constrain
quantitatively the contamination of our tracer populations by stars
that are too young to be dynamically relaxed, this contamination must
either be small or very similar across the different stellar magnitude
ranges. The similarity of the
power-law indices that we find for different tracers suggests that
they may indeed be representative for the actual underlying structure of
the old stars, which are expected to dominate the mass of the NSC.

As discussed in Paper\,I, the youngest stellar population is
  concentrated within $0.8\leq R\leq 12"$, or
  $0.03\leq R\leq 0.5$\,pc, of Sgr\,A*. It cannot be dynamically
  relaxed and the corresponding stars are therefore inadequate tracers
  of the putative cusp. We cannot directly measure the contamination
  of our SB profiles by pre-main sequence stars, but we can estimate it. Since
  the surface density of young stars is strongly peaked towards
  Sgr\,A*, this contamination is more severe at small $R$. As
  Figure\,12 in Paper\,I shows, the number density of pre-MS stars
  at $K_{s}\approx20$ is roughly two orders of magnitude
  below the one from the other stars at $R=2"$. We therefore conclude
  that contamination by pre-MS stars is not an issue for the SB
  profiles presented here.

As can be seen in the upper right panel of Fig.\,\ref{Fig:KLF}, the
population younger than 500\,Myr could contaminate significantly star
counts at magnitudes $17\lesssim K_{s}\lesssim19$. The importance of
this effect can currently not be well constrained because it depends
on the unknown distribution of stars in this age range. On the other
hand, the surface brightness measurements are dominated by older
stars. The fact that we observe similar surface densities and
brightnesses in Paper\,I is reassuring and suggests that contamination
effects are not severe.

\subsection{Optimised overall SB profile \label{sec:SB_opt}}

Mainly for illustrative purposes, we produced a 'best'  corrected
image by combining the corrected S13 and S27 wide-field images. The
images were matched via a least-squares minimisation of an additive
offset and a multiplicative scaling factor. We then measured the SB
profile again as in section\,\ref{sec:wide}, that is we also estimated
the uncertainty from inaccuracies in the sky background
subtraction. This case is very similar to the $K_{s}$ wide field
  data and a simple 2D power-law provides a good fit to the
  data in the range $R\leq1.0$ (blue dashed line in
  Fig.\,\ref{Fig:3D-simple}). The best-fit parameters are:
  $\Sigma_{0} = 15.6 \pm 0.1\,$mJy and $\Gamma=0.31\pm0.01$. We note
  that the fit deviates systematically from the data at $R\gtrsim25"$
  (see also case of $K_{s}$ wide field data shown in
  Fig.\,\ref{Fig:SB_WIDE}). This may suggest the necessity to use of a
  broken power-law. However,  this is not compelling once we take
  projection effects into account. A projected 3D simple power-law can
  fit the data well, as is shown by the straight orange line in
  Fig.\,\ref{Fig:3D-simple} and discussed in more detail in the
  following section.

\subsection{The 3D structure of the cluster \label{sec:nuker}}

Our observations provide us with the surface brightness, but we would
like to know the \emph{intrinsic} structure of the NSC. This is not a
trivial problem because it involves projection effects and
requires a fairly complete and accurate knowledge of the stellar
distribution on large scales both in and around the NSC. While we do
not yet possess very detailed knowledge -- in terms of high angular
resolution and multi-wavelength observations -- on the stellar population and
its distribution at large scales, we can use the results of previous work on the large
scale structure of the NSC combined with some basic or simplifying
assumptions (such as spherical symmetry) to provide an approximate,
general picture.

\begin{figure}[!htb]
\includegraphics[width=\columnwidth]{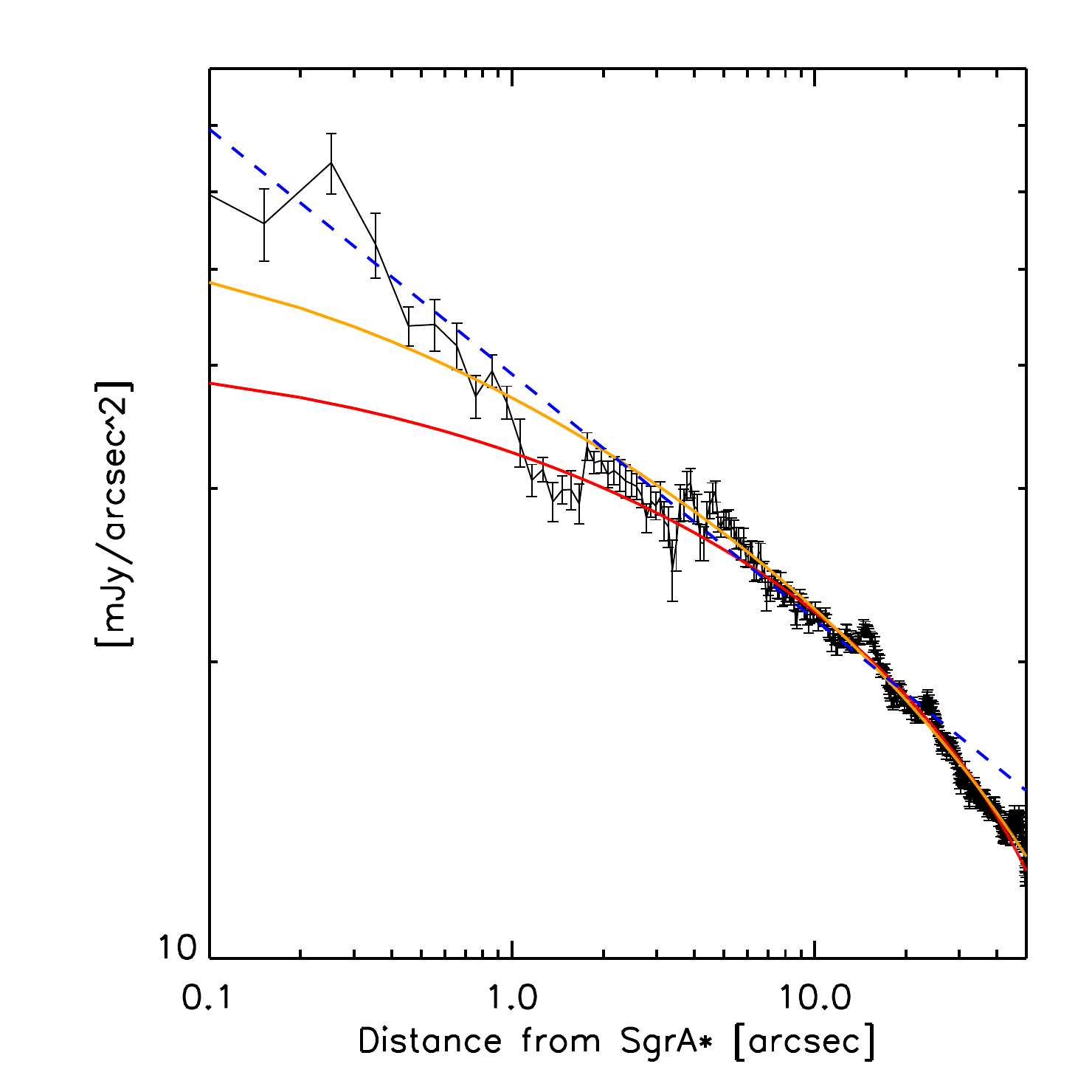}
\caption{\label{Fig:3D-simple} Black points: SB profile from the
  optimised $K_{s}$ wide plus $K_{s}$ S13 image, corrected for
  extinction and gas emission. The dashed blue line is a 2D simple
  power-law fit to the data at $R\leq25"$ (1\,pc). The straight lines
  are best fit projected 3D simple power-laws, assuming a simple
  model, where the stellar density drops to zero at outer cut-off
  radii of 10\,pc (red) and 20\,pc (orange).}
\end{figure}

  As a first, purely illustrative approximation to the
  problem, we assume a simple - and inaccurate - model, in which the
  intrinsic 3D structure of the cluster is described by a simple
  power-law with an outer cut-off, where the stellar density drops
  abruptly to zero. We use the combined $K_{s}$ wide field plus
  $K_{s}$ S13 data (see section\,\ref{sec:SB_opt}). They are shown in
  Fig.\,\ref{Fig:3D-simple} along with
  a best fit 2D simple power-law model and two best-fit projected 3D simple power-law SB
  models for two different outer cut-off radii. We note several
  points: (1) For a cluster of finite extent, the projected SB is
  \emph{not} given by a simple power-law. Instead, the SB continuously
  flattens towards small $R$.  The larger the outer cut-off radius,
  the closer the projected SB profile resembles a simple power-law
  (as can be expected). (2) For small clusters (or small assumed outer
  cut-off radii)), the change in the
  projected power law within $R\approx10"$ is so strong that a broken
  power-law may represent a better fit to the projected SB profile. In fact, such a broken
  power-law was frequently used in the past
  \citep[e.g.][]{Genzel:2003it,Schodel:2007tw,Buchholz:2009fk,Do:2009tg}. (3)
  In all cases, our simple toy model provides a surprisingly satisfactory
  fit to the data, with the best-fit value for the three-dimensional
  power-law index ranging between $\gamma=0.9-1.1$ (smaller value for the
  smaller cut-off radius).

\begin{table}
\centering
\caption{Best-fit model parameters for the Nuker fits to the SB profiles.}
\label{tab:models}
\begin{tabular}{l l l l l}
\hline
\hline
ID  & $r_{b}$ & $\gamma$ & $\beta$ & $\rho(r_{b})$ \\
& (pc)  & &  & (mJy\,arcsec$^{-3}$) \\
\hline
1$^{a}$ & $3.1\pm0.1$ & $1.11\pm0.01$ & $3.8\pm0.1$ & $0.029\pm0.002$ \\
2$^{b}$ & $2.9\pm0.2$ & $1.16\pm0.01$ & $2.9\pm0.1$ & $0.029\pm0.003$ \\
3$^{c}$ & $3.4\pm0.2$ & $1.17\pm0.02$ & $3.5\pm0.2$ &  $0.023\pm0.002$ \\
4$^{d}$ & $3.0\pm0.1$ & $1.12\pm0.01$ & $3.6\pm0.1$ & $0.029\pm0.001$ \\
5$^{e}$ & $3.0\pm0.1$ & $1.12\pm0.01$ & $3.6\pm0.1$ & $0.029\pm0.001$ \\
6$^{f}$ & $3.1\pm0.1$ & $1.12\pm0.01$ & $3.7\pm0.1$ & $0.029\pm0.002$ \\
7$^{g}$ & $3.2\pm0.2$ & $1.15\pm0.02$ & $3.3\pm0.1$ & $0.027\pm0.003$ \\
8$^{h}$ & $3.2\pm0.2$ & $1.15\pm0.02$ & $3.5\pm0.2$ & $0.026\pm0.003$ \\
9$^{i}$ & $3.2\pm0.2$ & $1.15\pm0.02$ & $3.3\pm0.1$ & $0.027\pm0.003$ \\
10$^{j}$ & $3.2\pm0.1$ & $1.14\pm0.01$ & $3.9\pm0.1$ & $0.028\pm0.002$ \\
11$^{k}$ & $3.4\pm0.2$ & $1.20\pm0.02$ & $3.4\pm0.2$ & $0.027\pm0.02$ \\
12$^{l}$ & $3.1\pm0.1$ & $1.12\pm0.01$ & $3.7\pm0.1$ & $0.029\pm0.002$ \\
13$^{m}$ & $3.0\pm0.1$ & $1.13\pm0.01$ & $3.6\pm0.1$ & $0.033\pm0.001$ \\
14$^{n}$ & $3.3\pm0.1$ & $1.09\pm0.01$ & $3.8\pm0.1$ & $0.023\pm0.001$ \\
16$^{o}$ & $2.3\pm0.1$ & $1.05\pm0.02$ & $3.0\pm0.0$ & $0.046\pm0.002$ \\
17$^{p}$ & $3.4\pm0.1$ & $1.14\pm0.01$ & $4.0\pm0.0$ & $0.024\pm0.001$ \\
19$^{q}$ & $3.6\pm0.2$ & $1.15\pm0.02$ & $3.6\pm0.2$ & $0.019\pm0.002$ \\
20$^{r}$ & $3.0\pm0.2$ & $1.15\pm0.02$ & $3.1\pm0.2$ & $0.028\pm0.003$ \\
\hline
\end{tabular}
\tablefoot{
\tablefoottext{a}{Data from \cite{Schodel:2014fk}, azimuthally
  averaged. Fore-/background emission model 2 of Table\,2 in \citet{Schodel:2014fk}. }\\
\tablefoottext{b}{Data from \cite{Schodel:2014fk} perpendicular to
  Galactic Plane.  Fore-/background emission model 2 of Table\,2 in \citet{Schodel:2014fk}.}\\
\tablefoottext{c}{Data from \cite{Schodel:2014fk} along
  Galactic Plane. Fore-/background emission model 2 of Table\,2 in \citet{Schodel:2014fk}.}\\
\tablefoottext{d}{Like $(a)$, but fore-/background emission model 5 of Table\,2 in \citet{Schodel:2014fk}.}\\
\tablefoottext{e}{Like $(a)$, but fore-/background emission model 4 of Table\,2 in \citet{Schodel:2014fk}.}\\
\tablefoottext{f}{Like $(d)$, but using only data at  $R\leq10\,$pc.}\\
\tablefoottext{g}{Data from \cite{Fritz:2016fj}. Fore-/background emission model 5 of Table\,2 in \citet{Schodel:2014fk}.}\\
\tablefoottext{h}{Like $(g)$, but fore-/background emission model 2 of Table\,2 in \citet{Schodel:2014fk}.}\\
\tablefoottext{i}{Like $(g)$, but fore-/background emission model 4 of Table\,2 in \citet{Schodel:2014fk}.}\\
\tablefoottext{j}{Like $(d)$, with lower integration boundary at $r= R  + 0.01$\,pc.}\\
\tablefoottext{k}{Like $(g)$, with lower integration boundary at $r= R  + 0.01$\,pc.}\\
\tablefoottext{l}{Like $(d)$, fitting only data at $R\leq10$\,pc.}\\
\tablefoottext{m}{Like $(d)$, with $\alpha=30$.}\\
\tablefoottext{n}{Like $(d)$, with $\alpha=5$.}\\
\tablefoottext{o}{Like $(d)$,with $\beta=3.0$ fixed.}\\
\tablefoottext{p}{Like $(d)$,with $\beta=4.0$ fixed.}\\
\tablefoottext{q}{Like $(g)$, but with $\alpha=5$. }\\
\tablefoottext{r}{Like $(g)$, fitting only data at $R\leq10$\,pc.}\\
}
\end{table}

Assuming an intrinsic simple power-law structure is, of course, an
oversimplification because a large body of previous studies of the
stellar density in the GC indicates that the nuclear cluster follows a
density of approximately $n(r)\propto r^{-2}$ outside of the central
parsec, with a steepening slope at larger distances \citep[see, e.g.
references and discussions in][]{Launhardt:2002nx,Schodel:2007tw,
  Schodel:2014fk,Fritz:2016fj}. A steepening density profile is also
required to avoid that the cluster mass diverges. Now we will explore
the consequences of a steeper density slope at larger distances on the
inferred three-dimensional power-law near Sgr\,A*.

To constrain the stellar distribution on scales of approximately 1 to
20\,pc, we use the data on the flux density of the NSC from
\citet{Schodel:2014fk} and \citet{Fritz:2016fj}. The former used
extinction-corrected Spitzer $4.5\,\mu$m surface brightness maps. The
latter used extinction corrected near-infrared data from NACO/VLT,
WFC/HST, and VISTA. Both data sets are not adequate to sample the
light density profile inside $R\approx 1\,$pc. The Spitzer data of
\citet{Schodel:2014fk} are of low-angular resolution and long
wavelength and completely dominated by a few bright stars and by
emission from the mini-spiral in the inner parsec. The data from
\citet{Fritz:2016fj} are, in principle, more suitable, but
are dominated by RC stars and brighter giants. As is well known and as
we confirm in Paper\,I, these stars show a core-like structure within
$R\leq0.3$\,pc from Sgr,A*.

 A caveat is that this previous work was focussed on significantly
  brighter stars than what we are examining in the present work. The
  data by \citet{Schodel:2014fk} and \citet{Fritz:2016fj} trace easily
  detectable stars and not the diffuse light density from very faint
  stars as analysed in this paper. Nevertheless, for simplicity -- and
  because we assume that it is a good approximation on large scales --
  we will assume that the distribution of all populations is
  described well by these data. A study of the density of different
  stellar populations throughout the nuclear cluster out to distances
  beyond a few parsecs is beyond the scope of this work and will be
  addressed in a later paper.  Again, we note that both in Paper\,I
  and in this work we find similar profiles for stellar components in
  significantly different brightness ranges, which supports our
  assumption that the individually detectable stars can be used as a good proxy for the
cluster shape on large scales.

To isolate the nuclear cluster from the emission of the nuclear disc
and Galactic Bulge, we used the S\'ersic models for the non-NSC
  emission listed in Table\,2 of \citet{Schodel:2014fk}. They were
  scaled to the data at $R\geq18$\,pc and subtracted from the data
  sets from \citet{Schodel:2014fk} and \citet{Fritz:2016fj},
  respectively.  We then scaled the latter data to our data in the
  ranges $1.5\,\mathrm{pc}\leq R\leq2.0\,$pc. At $R<1.5$\,pc, we
  used exclusively our data. We then applied a 3D ' Nuker' model and
  projected it onto the sky to fit the measured surface brightness. We
  use the Nuker model \citep{Lauer:1995fk} in the form of
  Equ.\,1 of \citet{Fritz:2016fj}:
\begin{equation}
\rho(r) = \rho(r_{b})2^{(\beta-\gamma)/\alpha}\left(\frac{r}{r_{b}}\right)^{-\gamma}\left[1+\left(\frac{r}{r_{b}}\right)^{\alpha}\right]^{(\gamma-\beta)/\alpha}.
\end{equation}

\begin{figure*}[!htb]
\includegraphics[width=\textwidth]{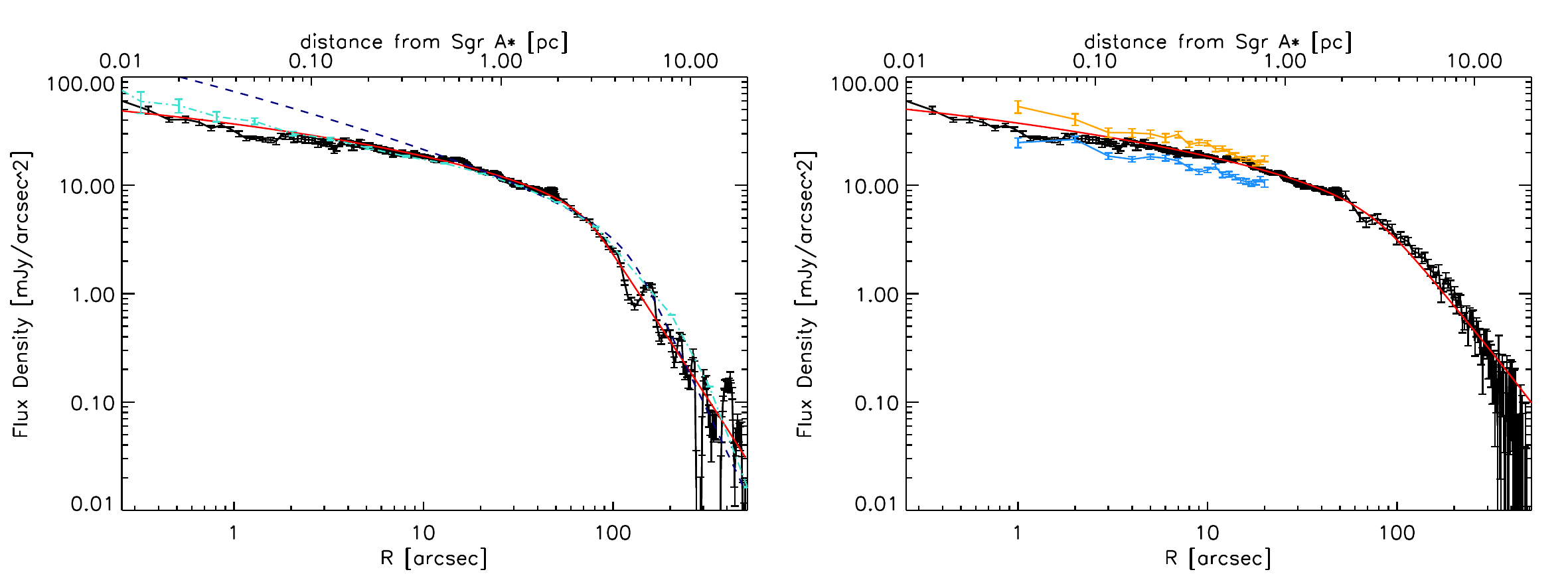}
\caption{\label{Fig:Nuker} Nuker model fits (red solid lines) to the diffuse
  SB in the GC. Left: The data at $R\geq1.5$\,pc are the azimuthally
  averaged, extinction corrected Spitzer $4.5\,\mu$m data from
  \citet{Schodel:2014fk} minus model to remove the contribution from
  components not part of the NSC. The dashed blue line is a fit
    with a forced inner slope of $\gamma=1.5$, corresponding to the
    lighter stars in a two-component Bahcall-Wolf cusp
    \citep{Bahcall:1977ys}. The dash-dotted turquoise line indicates
    the SB profile of the faint stars in the simulated cluster of
    Paper\,III.  Right: The data at $R\geq1.5$\,pc are the
  azimuthally averaged, extinction corrected near-infrared data from
  \citet{Fritz:2016fj} minus a model to remove the contribution from
  components not part of the NSC. The orange points and line show the
  stellar surface density data from Paper\,I for stars in the interval
  $K_{S}=17.5-18.5$ and blue for $K_{S}=16.5-17.5$. For better
  visualisation, the stellar surface densities have been scaled by
  constant, arbitrary factors.}
\end{figure*}

Here, $r$ is the 3D distance from Sgr\,A*, $r_{b}$ is the break
radius, $\rho$ is the 3D density, $\gamma$ is the exponent of the
inner and $\beta$ the one of the outer power-law, and $\alpha$ defines
the sharpness of the transition. We explicitly point out that the
  Nuker model was previously always used for 2D data, while we use it
  as a convenient mathematical model to describe the 3D shape of the cluster.
The density was then projected along the line of sight via an
integral:
\begin{equation}
\Sigma(R) = 2\int_{r}^{\infty}\frac{r \rho(r) dr}{\sqrt{r^{2}-R^{2}}}.
\end{equation}
For numerical reasons, to avoid a singularity, we could not integrate
down to $r=R$ and therefore set the minimum $r=R+0.001$\,pc. The
best-fit was found with the IDL MPFIT package
\citep{Markwardt:2009fk}. Uncertainties were re-scaled to a reduced
$\chi^{2}=1$. We fixed the parameter $\alpha=10$ and used only data at
$R\leq20$\,pc. Two of the fits, using the azimuthally averaged data of
\citet{Schodel:2014fk} and \citet{Fritz:2016fj} are shown in Fig.\,\ref{Fig:Nuker} (We note that the
plots corresponding to all fits performed by us have a  very similar appearance).

There are a number of obvious systematic uncertainties
related to this procedure. Our primary test of robustness is, of
course, the use of the completely independent data sets of \citet{Schodel:2014fk} and
\citet{Fritz:2016fj}. We then explored the parameter space by
repeating the fitting procedure for different cases:
\begin{itemize}
\item Fit with the azimuthally averaged data of \citet{Schodel:2014fk}
  as well as their profiles along the Galactic Plane and perpendicular
  to it, to examine the influence of the flattening of the nuclear
  cluster.
\item Flux offset due to non-NSC emission: We used different
    models from \citet{Schodel:2014fk} to estimate the fore- and
    background emission.
\item Fits with different settings for the minimum integration
  boundary ($r=0.001$ and $0.01$\,pc).
\item Fits for fixed different values of $\alpha=5,30$.
\item Fits to the entire data and fits limited to
  $R<10$\,pc to examine the influence of the fitting region.
\item Fits with a fixed parameter $\beta=3.0,4.0$.
\end{itemize}
Table\,\ref{tab:models} contains the best-fit parameters that we
obtained for the model-fits to different data and under different
assumptions and constraints. The
  $\chi^{2}$ values and the uncertainties of the different models and
  parameters are similar to each other. We can obtain an approximate,
  mean model for the nuclear cluster by taking the mean of each
  best-fit parameter and its standard deviation (not error of the
  mean; we do not include fixed parameters in these means):
  $r_{b} = 3.1\pm0.3$\,pc, $\gamma=1.13\pm0.03$, $\beta=3.5\pm0.3$,
  and $\rho(r_{b})=0.028\pm0.005$\,mJy\,arcsec$^{-3}$. It is important
  to note that there are covariances between these parameters. For
  example, the value of $\rho(r_{b})$ depends clearly on the value of
  $\beta$, with larger $\beta$ related to smaller
  $\rho(r_{b})$. Co-variance is also present between $r_{b}$ and
  $\beta$. On the other hand, the mean values
  are fairly well constrained and provide us with a good approximation
  of the overall 3D shape of the NSC. Finally, and most importantly
  with respect to the aim of this paper, the value of $\gamma$ is
  relatively tightly constrained and does not vary much between the
  different fits.

Theory predicts that the cusp
follows a power-law inside the break radius and that the latter
is on the order of the radius of influence of the black hole, which
has been found to be $\sim$$3-4$\,pc
\citep[e.g.][]{Alexander:2005fk,Feldmeier:2014kx,Fritz:2016fj,Feldmeier-Krause:2017rt},
consistent with The Nuker law  break radius determined here and in Paper\,I.

Here, we are most interested in the question of the existence of a
stellar cusp. As we can see, the three-dimensional power-law
  index $\gamma$
  can be determined robustly and is insensitive to the potential
  systematics that we have considered. Due to the finite structure of
  the cluster $\gamma$
  is not exactly equal to $\Gamma
  +
  1$.  The latter would only be valid for a simple power-law cluster
  with infinite extent. We assume that the systematic uncertainty of
  $\Gamma$,
  estimated to amount to $0.05$
  in section\,\ref{sec:Gamma}, applies also to $\gamma$.
  So, our best estimate for the 3D power-law index of the Milky Way's
  NSC in the innermost 1-2 parsecs is
  $\gamma=1.13\pm0.03_{model}\pm0.05_{sys}$,
  where the first uncertainty term is estimated from our Nuker model
  fits with different assumptions and constraints as listed in
  Table\,\ref{tab:models} and the second term is due to the
  measurements on different independent data sets as derived in
  section\,\ref{sec:Gamma}. Based on the value of $\gamma$
  and its uncertainty, we can rule out a flat core with high
  confidence.

An extensive analysis of the stellar number and flux surface density
in the GC was presented in \citet{Fritz:2016fj}. They fitted a
so-called $\gamma$-model and found that the radial structure of the
NSC in the innermost few $0.1$\,pc can be well described by a
power-law with index $\gamma=0.90\pm0.11$ for the stellar surface
density and $\gamma=0.76\pm0.08$.  for the flux density. These values
are flatter than what we have found here for the inner slope of the
cluster. The main difference between their work and our work is that
we focus on the diffuse emission of the faintest stellar population
while their measurements are dominated by giant stars.

In Paper\,I we show that the stars of $K_{S}\approx 17$ and $K_{S}\approx 18$ show a
projected surface density that is consistent with the one that we find
here for the diffuse light, while the giants
show a flattening inside a projected radius of $R\approx0.3$\,pc. In
Fig.\,\ref{Fig:Nuker} we overplot the stellar surface number densities
onto the plot of the surface brightness density of the diffuse light.

We find, however, larger values of $\gamma$ in Paper\,I. For the
$K_{s}\approx18$ stars we find $\gamma=1.41\pm0.06\pm0.1_{sys}$. This discrepancy
may indicate certain biases related to the different methods. For
example, we may have underestimated the dynamically unrelaxed stars
may contaminate the star counts, as discussed in Paper\,I or we may  over-estimated incompleteness due to
crowding. Alternatively, we may have over-corrected the emission from
gas and dust in this work, there may be a bias from the sky background
subtraction resulting from an observational setup that was not
optimised for measuring the unresolved, diffuse emission, or the
different values reflect uncertainties in the scaled matching of our
measurements and literature data at $R>1.5$\,pc. There may
also be other systematic effects at play that we have not
considered. We note, however, that both values for $\gamma$ exclude a
flat, core-like profile with high confidence. Their difference
  can provide us with a robust estimate of the true systematic
  uncertainty of both values, which may thus be on the order of
  $\Delta\gamma_{sys}=0.15$.

In Paper III (Baumgardt et al.\ arXiv:1701.03818) we compare the
measurements to N-body simulations and confirm the consistency between
measurements and theory. The probably best explanation for the
flatness of the observed cusp is mass segregation between stars of
different masses in the inner parts of the nuclear cluster, which
flattens the density profile of bright stars away from the
$\gamma=1.75$ prediction of \citet{Bahcall:1976vn}.  In addition, due
to repeated star formation and/or cluster infall not all the stars in
the nuclear cluster may be old enough to be fully dynamically relaxed,
which could cause a further modification of the central slope.  Most
models created so far assumed clusters with a single age stellar
population that evolved for many relaxation times. The NSC of the
Milky Way, on the other hand, contains stellar population of different
ages \citep[see, e.g.][]{Pfuhl:2011uq}. Also, the NSC may have had
less than a Hubble time for two body relaxation processes to work, so
it may not be fully relaxed. Paper\,III presents more elaborate
theoretical models, based on direct N-body simulations and explicit
consideration of the
star formation history of the NSC \citep[modelled according to the one
derived in][]{Pfuhl:2011uq}, that provide
results consistent with our data. We believe that the relative
flatness of the cusp is the reason why it has eluded any clear
confirmation for decades.

As concerns the value of $\beta$, which describes the density decrease
at distances $r>>r_{b}$, we find in Paper\,I a value
$\beta_{resolved}=3.4\pm0.3$, which agrees well with the value derived
here and in earlier work
on the large-scale structure of the NSC  \citep[see introduction and
references in][]{Schodel:2007tw}. As a final note,
the data used here to constrain the cluster structure at large $R$
reflect a much brighter tracer population than the stars that dominate
the diffuse emission from unresolved stars in our NACO images.

\subsection{Density of stars near Sgr\,A*, enclosed stellar mass}

For observational purposes, it is of great interest to obtain a rough
estimate of the surface number density of unresolved stars at
$R=0.25"$ ($R=0.01$\,pc). On the one hand, the results from Paper\,I
show that the surface number density of stars at
$17.5\leq K_{s} \leq 18.5$ is about 20\,arcsec$^{-2}$ at
$R=0.25"$. Applying this normalisation to the model KLF from
section\,\ref{sec:masses}, this corresponds to 80 stars arcsec$^{-2}$
in the interval $18.5\leq K_{s} \leq 19.5$ and 370 stars arcsec$^{-2}$
in the interval $19.5\leq K_{s} \leq 20.5$.  If we use the surface
flux density derived in this work, on the other hand, we obtain
somewhat different, but consistent, values. The extinction-corrected
surface flux density estimated at $R=0.01$\,pc is about
50\,mJy\,arcsec$^{-2}$, which results, for the same model KLF,
densities of 64 stars arcsec$^{-2}$ in the interval
$18.5\leq K_{s} \leq 19.5$ and 300 stars arcsec$^{-2}$ in the interval
$19.5\leq K_{s} \leq 20.5$. The comparison between the numbers of
  faint stars obtained by these two estimates points to an uncertainty
  of about $20\% $. An additional source of uncertainty, also on the
  order of $20\%$, results from the exact normalisation of the KLF.
Here we assumed that the diffuse flux is dominated by stars
$19\leq K_{s} \leq 22$. NIR cameras at the next generation of
extremely large telescopes, such as MICADO/E-ELT
\citep{2010Msngr.140...32D}, will have angular resolutions of
$\lesssim10$\,mas FWHM, and thus be able to resolve surface number
densities on the order of 1000 stars arcsec$^{-2}$ (the actual
performance will depend on the dynamical range and luminosity function
of the observed field). Hence, the future generation of ground-based,
AO-assisted telescopes will be able to observe the stellar cusp around
Sgr\,A* directly, down to about one solar mass stars. The high stellar
surface density is encouraging for interferometric observations of the
immediate environment of Sgr\,A* with an instrument such as
GRAVITY/VLTI \citep{Eisenhauer:2011vn}, if it can reach the required
high sensitivity.

\begin{table}
\centering
\caption{Stellar mass densities near Sgr\,A* and total stellar mass
  within $r=1\,pc$.}
\label{tab:massdensity}
\begin{tabular}{l l l l l}
\hline
\hline
ID  & $\rho(1\,\mathrm{pc})$ & $\rho(0.1\,\mathrm{pc})$ & $\rho(0.01\,\mathrm{pc})$  & $m_{stellar}(1\,\mathrm{pc})$\\
& (M$_{\odot}$\,pc$^{-3}$)  & (M$_{\odot}$\,pc$^{-3}$)  & (M$_{\odot}$\,pc$^{-3}$) & M$_{\odot}$ \\
\hline
1$^{a}$ & $1.5\times10^{5}$ & $2.0\times10^{6}$ & $2.7\times10^{7}$ & $1.0\times10^{6}$\\
2$^{b}$ & $1.8\times10^{5}$ & $2.4\times10^{6}$ & $3.2\times10^{7}$ & $1.2\times10^{6}$\\
3$^{c}$ & $1.6\times10^{5}$ & $2.2\times10^{6}$ & $3.0\times10^{7}$ & $1.1\times10^{6}$\\
4$^{d}$ & $1.2\times10^{5}$ & $1.6\times10^{6}$ & $2.2\times10^{7}$ & $0.8\times10^{6}$\\
\hline
\end{tabular}
\tablefoot{
\tablefoottext{a}{Normalisation to a total cluster mass of $2.5\times10^{7}$\,M$_{\odot}$ \citep{Schodel:2014fk}.}\\
\tablefoottext{b}{Normalisation to a mass of  $1.4\times10^{7}$\,M$_{\odot}$
within $4.2$\,pc of Sgr\,A* \citep{Feldmeier:2014kx}.}\\
\tablefoottext{c}{Normalisation to a mass of $1.1\times10^{6}$\,M$_{\odot}$
within $1$\,pc of Sgr\,A* \citep{Schodel:2009zr}.}\\
\tablefoottext{d}{Normalisation to a mass of $8.94\times10^{6}$\,M$_{\odot}$
within $3.9$\,pc of Sgr\,A* \citep{Chatzopoulos:2015yu}.}\\
}
\end{table}

Another value of interest is the mass density near Sgr\,A* and the
total enclosed mass within 1\,pc of Sgr\,A*. Using our best-fit
Nuker-law parameters, we have computed the mass density at distances
of $r=1,0.1$, and $0.01$\,pc from Sgr\,A*, using five different
normalisations of the enclosed mass, four of them dynamical
\citep{Schodel:2009zr,Feldmeier:2014kx,Chatzopoulos:2015yu,Fritz:2016fj}
and one of them based on mass-to-light ratio \citep{Schodel:2014fk}.
The values are listed in Table\,\ref{tab:massdensity} and agree within
factors of less than two.

Densities in excess of a few $10^{7}$\,M$_{\odot}$\,pc$^{-3}$ are
reached at $r<0.01$\,pc of Sgr\,A*, which corresponds roughly to the
apo-centre of the orbit of the short-period star S2/S0-2
\citep[e.g.][]{Boehle:2016zr}. This is comparable to what has been
inferred by some models for the central density of Omega Centauri
\citep{Noyola:2008fk}. We note that $0.01$\,pc correspond to about
$0.25"$ or a few resolution elements of a 10m-class telescope in the
NIR at the distance of the GC. In spite of this high density, the
small volume implies that this corresponds to only
$180\pm30$\,M$_{\odot}$ (taking the mean and standard deviation
of the estimates resulting from the different normalisations).

From the different values given in Table\,\ref{tab:massdensity}, we
estimate a total stellar mass within $r=0.1$\,pc of Sgr\,A* of about
$1.3\pm0.1\times10^{4}\,$M$_{\odot}$, smaller than, but of the same
order of magnitude as, the value given by \citet{Yusef-Zadeh:2012pd}.
The total mass within $r=1$\,pc of Sgr\,A* is
  $1.0\pm0.1\times10^{6}\,$M$_{\odot}$ and within $r=3$\,pc of Sgr\,A*
  is $7.8\pm0.6\times10^{6}\,$M$_{\odot}$, roughly twice the mass of
  Sgr\,A*. As a note of caution, we remind the reader here that the
Nuker model assumed here for the NSC does not take into account the
mass from the nuclear bulge or other stellar components that do not
form part of the NSC, but may overlap with it. Therefore, our model
will under-estimate the real mass enclosed at large~$r$.

We point out that here we assume a constant mass-to-light ratio
throughout the NSC. This may result in an under-estimation of the
enclosed mass of the NSC at small radii.  Theoretical considerations
and simulations predict an accumulation of stellar-mass black holes in
an invisible, steep ($\gamma\approx-1.75$) cusp around Sgr\,A*
\citep[e.g.][]{Morris:1993ve,Merritt:2006ys,Alexander:2009gd,Preto:2010kx}.
This cusp is actually steeper when one considers realistic number
fractions for the stellar population, which leads to a more efficient
segregation of the masses. In particular, \citet{Alexander:2009gd},
\citet{Preto:2010kx}, and \citet{Amaro-Seoane:2011qv} find in their
models that the cusp for their 'heavy' stars, the precursors of
stellar-mass black holes, build up a cusp with $\gamma\approx
-2$.
They refer to this finding as ``strong mass segregation''.  Depending
on the properties of this putative black hole cusp, the enclosed mass
at small distances from Sgr\,A* may be significantly higher than the
estimates provided here. The most recent constraint on the extended
mass within $0.01$\,pc of Sgr\,A* from the orbital analysis of
individual stars is that it must be less than
$1.3\times10^{5}$\,M$_{\odot}$ \citep{Boehle:2016zr}. Hence, the mass
density estimated here can be easily accommodated by current
dynamical analyses.


\section{Conclusions}

This paper presents the radial surface brightness profile of the
diffuse emission in high angular resolution, point source-subtracted
images of the GC. After taking into account the contamination of the
diffuse light by line emission from  gas and dust in the mini-spiral, we argue that the
diffuse emission arises from a faint, unresolved stellar population
with magnitudes of $K_{S}= 19-22$. This corresponds to main sequence
stars or sub-giant stars with masses of about
$0.8-1.5\,M_{\odot}$. These stars can live long enough
on the main sequence to be
dynamically relaxed and thus to serve as a tracer population for a
stellar cusp around the central black hole of the Milky Way.

We find that the projected surface brightness profile can be
  fitted well by a power-law slope with an index of
  $\Gamma_{in}=0.26\pm0.02_{stat}\pm0.05_{sys}$ at $R<0.5$\,pc. This
  value is smaller than, but consistent with what we find for the
  stellar surface number density of $K_{s}\approx17$ and
  $K_{s}\approx18$ (observed magnitude) stars in Paper\,I. An
  important caveat is that we cannot directly determine which kind of
  stars we are observing and the contamination of the star counts by
  young, dynamically unrelaxed stars may be high, as discussed in
  Paper\,I. However, the fact that the work in this paper and in
  Paper\,I use different methodologies, but arrive a similar results,
  gives us confidence in our results.

Translating these results into an intrinsic, three-dimensional
description of the cluster is not trivial, but by using previous
studies of the cluster morphology on large scales as constraints,
along with a spherical approximation, we find that the cluster can be
described well by a three-dimensional Nuker law within about
20\,pc of the central black hole. According to our models, the break
radius is $3.1\pm0.3$\,pc contains about a stellar mass of twice the
mass of Sgr\,A* and thus coincides with the radius of influence of the
black hole \citep[e.g.][]{Alexander:2005fk}. The three-dimensional
density inside of the break radius follows a power law with an
exponent $\gamma_{in}=1.13\pm0.03_{model}\pm0.05_{sys}$. A core-like
distribution of the faint stars can thus be firmly excluded. From a
comparison between the results for the faint, unresolved stellar
population and the faint resolved population (Paper\,I), we
suggest that a robust range for the power-law index of the cusp is
$\gamma=1.1-1.4$.

An underlying assumption of our work is that the faint emission arises
indeed mostly from stars old enough to be dynamically relaxed. A
possible source of concern could be contamination by pre-main sequence
stars in the region of the few million year-old starburst within
$R=0.5$\,pc of Sgr\,A*. Our analysis, in Paper\,I, of the KLF of
  the stars in the inner parsec, shows that this possibility is
rather  unlikely.

The stellar cusp identified in this work and in Paper\,I is
flatter than the one predicted for single-mass stars
around a massive central black hole $\gamma_{theor}=1.75$, or for
low-mass stars in a cluster composed of two mass groups
($\gamma_{theor}=1.5$). In contrast to the simplifying assumptions of
previous theoretical work, the nuclear cluster at the GC has undergone
multiple epochs of star formation and/or cluster infall.  Thus, not
all the stars may be old enough to be fully dynamically relaxed.  As
we will elaborate in Paper III, our observations nicely agree
with the detailed, direct-summation Nbody simulations.In
Paper\,III we compare the measurements to N-body simulations and
confirm the consistency between measurements and theory.

The flatness of the cusp is one of the main reasons why it may have eluded
detection so far.  The second reason is that the giant stars brighter
than $K_{S}\approx16$ dominated all previous attempts at determining
the NSC's structure. However, these stars show a core-like profile
in projection within $R\approx0.3$\,pc (see Paper\,I and discussion and references
therein).

We summarise our conclusions here:
\begin{enumerate}
  \item Our study of the diffuse stellar light around Sgr\,A* confirms the existence of a simple power-law cusp
    around Sgr\,A*, with a 3D power-law index $\gamma\approx1.13\pm0.03_{model}\pm0.05_{sys}$.
  \item The cusp is shallower than what is predicted by theory.
   \item The existence of a cusp in our Galaxy supports the existence
    of stellar cusps in other, similar systems that are composed of a
    nuclear cluster and a massive black hole.
   \item The existence of stellar cusps is an important prerequisite
     for the observation of EMRIs with
     gravitational wave detectors.
   \item The bright giants and the Red Clump stars at the GC do not
     show the same distribution as the fainter stars. Either the
     bright giants are, on average, younger than the fainter stars and
     are not yet dynamically sufficiently well relaxed, or some
     mechanism has altered the appearance of this population:
       Possibly, the envelope of giants were removed by colliding
     with the fragmenting gas disc at the GC which later turned into
     the observed stellar disc of young, massive stars
     \citep{Amaro-Seoane:2014fk}.
\end{enumerate}

Future research needs to be done to refine our understanding of the
cusp at the GC. On the observational side, we need to infer robust
data on the large-scale two-dimensional distribution of stars out to
about 10 \,pc from Sgr\,A* with high sensitivity and angular
resolution. We will then be able to reconstruct the intrinsic
three-dimensional profile of the cluster. The next step will then be
an accurate determination of the different types of faint stars near
Sgr\,A* (e.g.: Which ones are pre-MS stars?) in order to understand
the age structure of the nuclear star cluster. At least some of this
future work can only be done with a 30m-class telescope. Observations
with the next generation of telescopes can test the predictions on
stellar number densities from our work.

\begin{acknowledgements}
  The research leading to these results has received funding from the
  European Research Council under the European Union's Seventh
  Framework Programme (FP7/2007-2013) / ERC grant agreement
  n$^{\circ}$ [614922]. PAS acknowledges support from the Ram{\'o}n y
  Cajal Programme of the Spanish Ministerio de Economía, Industria y
  Competitividad. This work has been partially supported by the CAS
  President's International Fellowship Initiative. FNL acknowledges
  financial support from a predoctoral contract of the Spanish
  Ministerio de Educación, Cultura y Deporte, code FPU14/01700.  This
  work is based on observations made with ESO Telescopes at the La
  Silla Paranal Observatory under programmes IDs 083.B-0390,
  183.B-0100 and 089.B-0162. We thank the staff of ESO for their great
  efforts and helpfulness. We thank Tobias Fritz for detailed and
  valuable comments.
\end{acknowledgements}

%
%

\begin{appendix} 

\section{Photometric accuracy and recovery of diffuse light with
  StarFinder \label{app:photo}}

In this section we explore two issues via simulations of the Galactic Centre: (1) Photometric
accuracy and point-source residuals when the PSF varies across the
field due to anisoplanatic effects. (2) The capability of recovering
the diffuse light with Starfinder in a GC-like environment and with a
spatially variable PSF. As a test case we use observations of NACO
through the Br$\gamma$ filter, where the diffuse background is
particularly high and variable due to the strong line emission from
the minispiral.

The simulated images are based on the Br$\gamma$ observations
described in section\,\ref{sec:data}. We used stellar sources detected
down to $K_{s}=18$, where the star counts are reasonably complete
across the field and source detection is highly reliable. The guide
star PSF was extracted from IRS\,7 in the original Br$\gamma$ data,
after having repaired the saturated core of IRS\,7. To simulate the
variation of the PSF across the field, we modelled the loss of Strehl
and elongation of the PSFs via convolution with Gaussian kernels. The
latter are chosen as being elongated along the line connecting any
given star to the guide star, a typical manifestation of
anisoplanatic effects. The FWHM of the Gaussians along these lines
grows by $0.027''$ for every $10''$ distance from IRS\,7 and by
$0.008"$ in the perpendicular direction. In this way we obtain a
simulated image that appears similar to the original image, albeit
with a somewhat stronger anisoplanatic effect, which is good because it
means that we are carrying out our simulations for a conservative test
case. We use a FWHM of the PSF of about $0.08"$ for the guide star. At
a distance of $20"$ from the guide star, the PSF has a FWHM of about
$0.1"$ along the line connecting it with the guide star.

Finally, we added readout and photon
noise (both from sources and from the sky). We then carried out runs
of {\it StarFinder} both with a constant and with a variable PSF, the
latter as described in section\,\ref{sec:data}.

We simulated images with a flat, zero background, with a complex
background by including the minispiral, and with a complex background
that includes the minispiral and an additional diffuse power-law
component. As described in section\,\ref{sec:data}, we do not fit the
background with {\it StarFinder}, that is, the keywords $BACK\_BOX$ and
$ESTIMATE\_BG$ are set to zero. Instead, we determined the diffuse
light directly from the point-source-subtracted images. As a side
note, we point out that throughout this paper we use the terms
background and diffuse emission in an equivalent way.

In all simulated images, we first repaired the core of the PSF of the
brightest star, IRS\,7. Although it was not saturated in the simulated
images, of course, it is saturated in the real data and this step is
necessary because if forms an integral part of the data reduction. It
will lead to a slight broadening of the guide star's PSF because its
core is replaced by the median of the cores of nearby stars, that have
a somewhat lower Strehl. We note that we wrote our own code for
repairing the core of IRS\,7 because the native {\it StarFinder} code
for this purpose, REPAIR\_SATURATED.PRO will only work
accurately if the complete PSF is known a priori. However, this is not
the case here, where we actually use the brightest, saturated star to
estimate the broad, extended wings of the PSF. We mention this problem
here because it may arise in most similar situations where a user
wants to repair the cores of saturated stars with {\it
  StarFinder}. The key is that, while {\it StarFinder} only applies a
multiplicative scaling factor, one must also use an additive offset
when fitting the core to the saturated star.

When we apply PSF fitting with a variable PSF, we sub-divide the
simulated image into overlapping square fields of $10.8"$ size on a
side, that we call 'sub-images'. A local PSF is estimated from the
brightest, isolated stars in each field. Subsequently, the wings from
the bright guide star are fitted to this local PSF and PSF fitting is
performed on each sub-image. The sub-images overlap by half of their
size. When recomposing the point-source-subtracted images, the
borders of the sub-images (about $2"$ width) are removed and the
remaining overlapping areas are averaged. In this way we can create a
homogeneous residual image.

\subsection{Variable PSF and constant zero background}

\begin{figure*}[!htb]
\includegraphics[width=\textwidth]{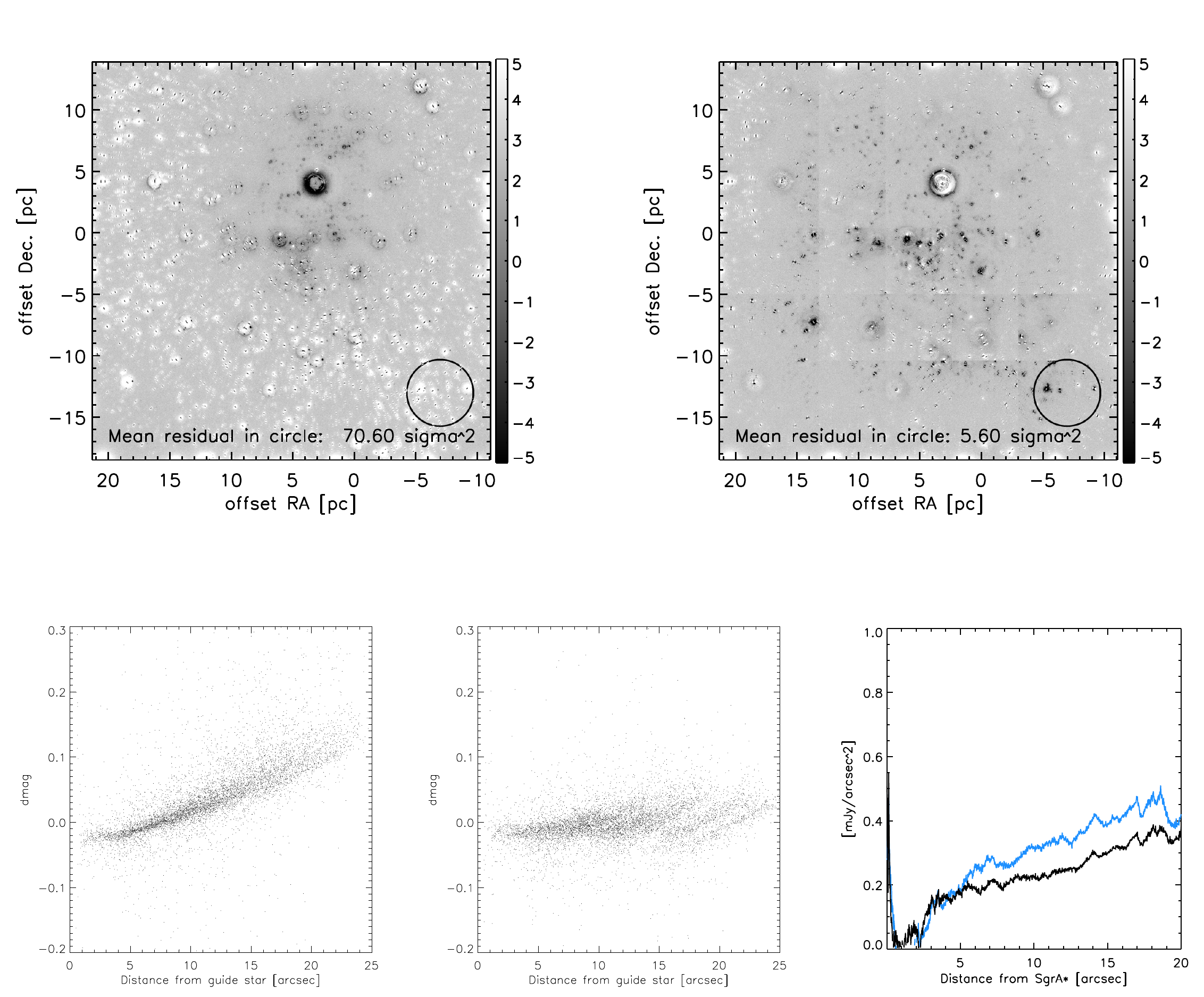}
\caption{\label{Fig:app1} PSF fitting test with a variable PSF and a
  constant background. Upper left: Point-source subtracted image after
  use of a single, constant PSF. The circle in the lower
  right shows a region in which we measured the sum of the squared
  residual, which is $70.6\,\sigma^{2}$ for this region. Upper right:
  Like upper left, but after using a variable PSF. The sum of the
  squared residual in the circle is $5.6\,\sigma^{2}$. The grey scales are expressed in terms
  of $\sigma$ deviations from the noise image.  Bottom left:
  Differences between the measured magnitudes of stars and their input
  magnitudes when a single, constant PSF is used. Bottom centre:
  Differences between the measured magnitudes of stars and their input
  magnitudes when a variable PSF is used. Bottom right: Plot of
  background, after point-source subtraction, as a function of
  distance from Sgr\,A*. The background is the median in rings around
  Sgr\,A*. The blue data are for the case of a single PSF and the
  black data for the case of a variable PSF. }
\end{figure*}

Our first test is performed with a constant background of value
zero. Several images and plots that evaluate this test quantitatively
are shown in Fig.\,\ref{Fig:app1}.

We can see that the residuals are significant and systematic in case
of using only a single, constant PSF. They are significantly smaller
and more constant across the field when we use a variable
PSF. Quantitatively, this effect can be seen nicely in the plot of the
differences between measured and input magnitudes for the stars. They
show a systematic trend with distance from the guide star in case of
use of a single PSF. With a variable PSF, some local systematics
appear (as expected because we do not model the PSF for each
position), but they are far smaller. In general, systematic
photometric uncertainties due to PSF variability are on the order of
just a few $0.01$\,mag when we use a variable PSF. As concerns the
measured background, after point-source subtraction, it is close to
zero, but has a small, positive bias that increases with distance from
the guide star.  This trend can possibly be partially explained by the
fact that measurements become less accurate towards the image edges
because we cannot minimise uncertainties by multiple measurements in
overlapping fields near the edges and because the potential PSF
reference sources are fewer and fainter towards the image
edges. Variable PSF fitting performs better in background recovery
than constant PSF fitting. We also note that there is a small dip of
the recovered background close to the position of Sgr\,A*. We believe
that this is due to the strong concentration of bright stars there. It
appears that the broadening of the PSF caused by the necessary
superposition of several reference stars leads to negative residuals
close to bright stars. In any case, the positive residual is very
small. It is, in the worst case, not more than a few percent of the
surface brightness of the mini-spiral or diffuse stellar emission that
we analyse in this work. We thus conclude that it is safe to ignore it.  We also conclude
that using a variable PSF is superior to using a constant PSF and that
a constant flat background $\gtrsim1.0$\,mJy\,arcsec$^{-2}$ can be
accurately recovered by our method.

\subsection{Variable PSF plus complex diffuse emission from gas}

\begin{figure*}[!htb]
\includegraphics[width=\textwidth]{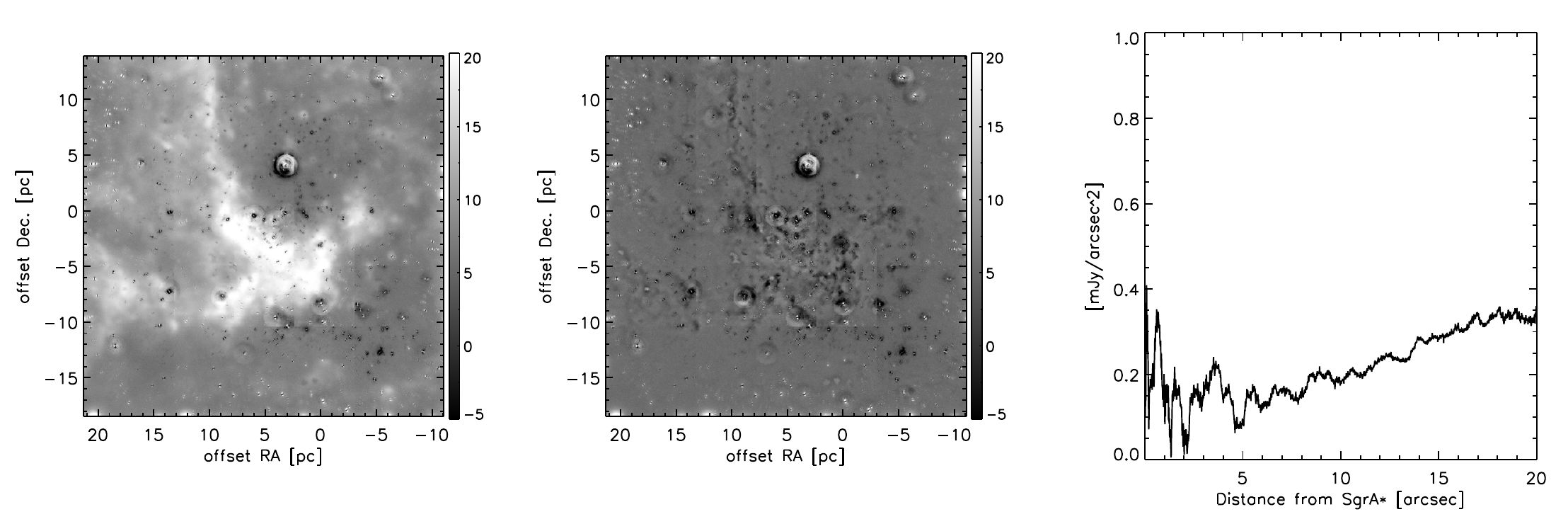}
\caption{\label{Fig:app2} PSF fitting test with a variable PSF and  a
  complex, diffuse background. Left: Point-source subtracted image after
  use of a variable PSF.  Middle: Residual image, after subtracting the
  input distribution of diffuse emission. Right: Background
  as function of distance from Sgr\,A* as measured in the residual
  image. The grey scales are expressed in terms
  of $\sigma$ deviations from the noise image. }
\end{figure*}

To model highly complex diffuse emission we used the HST
Paschen\,$\alpha$ image of the minispiral, transformed it to the frame
of the NACO Brackett-$\gamma$ image, and scaled its flux
accordingly. Some smoothing was applied to mitigate the effects of
interpolation. Then we proceeded as described above. In
Fig.\,\ref{Fig:app2} we show the simulated image after fitting and
subtracting the point-sources and the residual after, additionally,
subtracting the input gas emission. Finally, we show the plot of
residual light density as a function of distance from Sgr\,A*. It is
close to zero at all distances, very similar as in the case of a flat,
zero background. We conclude that our variable PSF fitting with {\it
  StarFinder} can reproduce very well the details of complex diffuse
emission and that the residual light can be reproduced accurately after
the complex diffuse emission is removed. Here, we remind the reader
again that we do not fit the diffuse emission with {\it
  StarFinder}. We just fit and subtract the point-sources.

\subsection{Variable PSF plus gas and  power-law cusp}

\begin{figure*}[!htb]
\includegraphics[width=\textwidth]{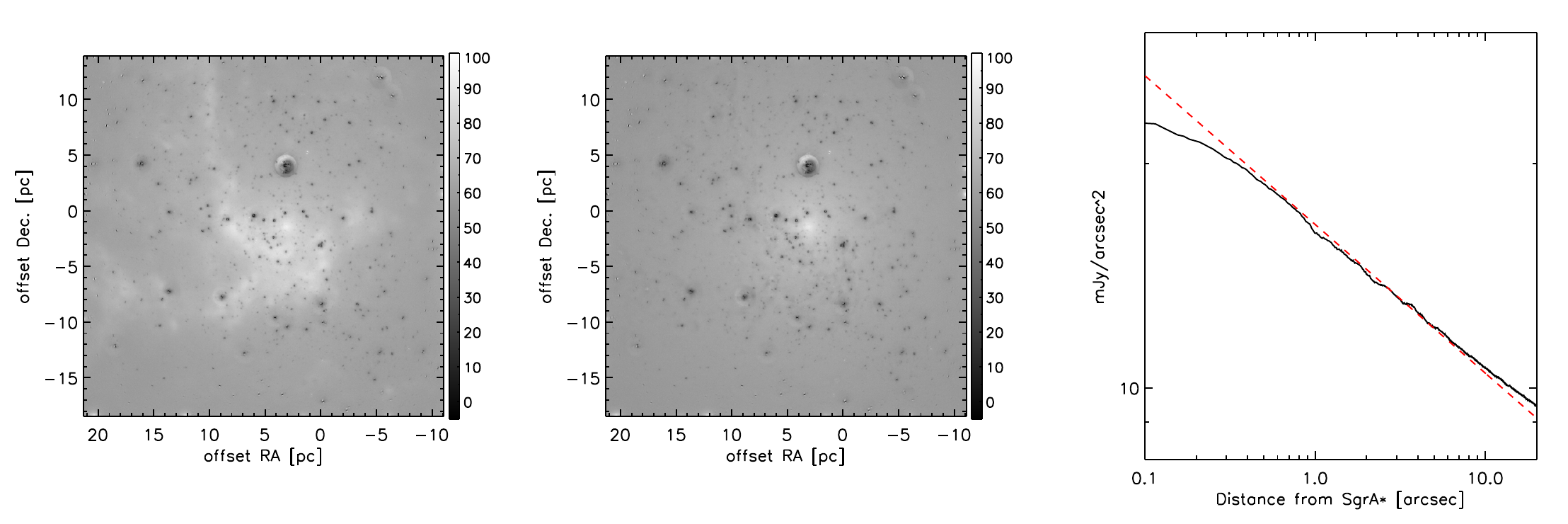}
\caption{\label{Fig:app3} PSF fitting test with a variable PSF,
  complex, extended emission from  gas/dust, and diffuse emission
  from a population of unresolved faint stars. Left: Point-source
  subtracted image after use of a variable PSF. The grey scale is
  expressed in terms of $\sigma$ deviations from the noise image.
  Middle: Residual image, after subtracting the input distribution of
  diffuse emission. Right: Background as function of distance from
  Sgr\,A* as measured in the residual image. The dashed red line is
  the power-law cusp used as input. The grey scales are expressed in
  terms of $\sigma$ deviations from the noise image.  We note that the
  scales are different from the ones used in in Fig.\,\ref{Fig:app2}}
\end{figure*}

Finally, we added a power-law cusp from faint, diffuse stellar
emission to the simulated image, proceeding as described in the
previous sections. The cusp was simulated as a pure power-law with a
2D exponent of $\Gamma=-0.2$, a scale radius of $R_{0}=12.5"$, and a
flux density of 10\,mJy\,arcsec$^{2}$ at $R_{0}$.  In
Fig.\,\ref{Fig:app3} we show the simulated image after fitting and
subtracting the point-sources and the residual after, additionally,
subtracting the gas emission. Finally, we show the plot of residual
light density as a function of distance from Sgr\,A* with the input
cusp model over-plotted. The recovered power-law cusp is almost
identical to the input model, with minor deviations only near the edge
of the field and near Sgr\,A*.

\subsection{Variable PSF: Real data }

\begin{figure*}[!htb]
\includegraphics[width=\textwidth]{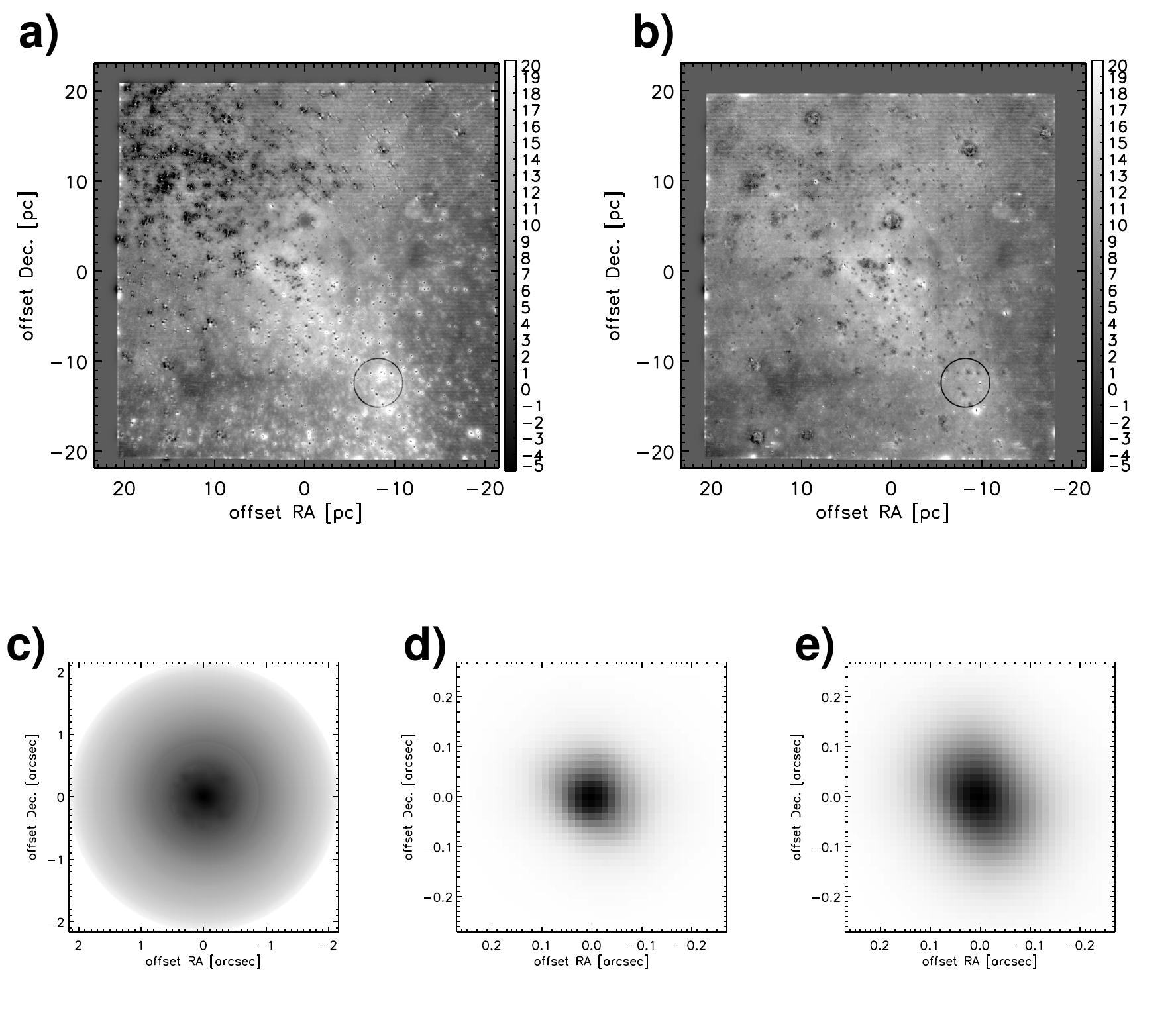}
\caption{\label{Fig:app4} PSF fitting test with a variable PSF on the
  H-band image used in this work. a) Point-source-subtracted image
  after use of a single, constant PSF. The circle in the lower
  right shows a region in which we measured the sum of the squared
  residual, which is $131\,\sigma^{2}$. b) Point-source-subtracted image
  after use of a variable PSF. The circle in the lower
  right shows a region in which we measured the sum of the squared
  residual, which is $41\,\sigma^{2}$. c) Full extent of the PSF. d)
  Zoom onto core of PSF near image centre. e) Zoom onto core of PSF
  near the lower right corner. All grey scale are logarithmic.}
\end{figure*}

Finally, we will take a closer look at the performance of our
methodology with real data. For this purpose we use the $H-$band image
used in this work. The point-source-subtracted images for a constant
PSF (panel a)) and use of a variable PSF (panel  b)) are shown in
Fig.\,\ref{Fig:app4}. Significant systematic residuals related to
point-sources can be seen in case of the constant PSF. Moreover, those
residuals vary strongly with position in the field. We point also out
that the real data show a different residual pattern than the
simulated data. In particular, the residuals look less symmetric than
in case of the simulated data. We believe that this can probably be
explained by time-variable AO performance because the final mosaic
image is the result of observations of four different
pointings. Variable AO perfomance is a frequent feature of AO
instruments and is mostly related to changes in the atmospheric
seeing. We did, however, not further investigate this effect here
because it would go far beyond the purpose of this paper.

The point-source-subtracted image after using a variable PSF shows
smaller residuals that are more homogeneous across the field. The
total squared residuals are significantly lower than in case of a
constant PSF. We note that the residuals here do still include
diffuse emission from gas and unresolved stars.

Panel c) of Fig.\,\ref{Fig:app4} shows the full extent of the PSF
halo. It can be seen that it extends out to almost $2''$ from the the
centre of the PSF. Panels c) and d) show zooms onto the cores of
locally estimated PSFs. It can be seen that the PSF core near the image
edge, at roughly $15"$ from the guide star, is elongated compared to
the PSF core near the image centre.

The conclusion of this appendix is that our variable PSF fitting  with
{\it StarFinder} with subsequent point-source-subtraction is well
suited to recover the complex diffuse emission from gas/dust and,
after removal of the latter, to measure the diffuse distribution from
a faint, unresolved stellar population at the GC.

\section{Systematic errors of the 2D fit of the SB profile\label{sec:app2}}

In this section we examine several potential sources of systematic
errors in the power law plus scaled gas emission fits to the SFP
profiles of the $K_{s}$ wide field image. The results are readily
applicable to the images in other filters.

\subsection{Sky subtraction}

Our experiments with the data suggest that the strongest systematic
effect can arise from an unknown additive offset of the diffuse
emission. Although the images were sky-subtracted -- the sky
background in the corresponding filters was measured on a dark cloud
at a few arcminutes offset -- there is some uncertainty related to
this procedure: There was only one sky measurement done for the
approximately one hour-long observations.  Hence, the sky background
may have varied.  Also, we are interested in the SB profile of the
nuclear star cluster, but there may be diffuse flux contributions from
other structures, such as the nuclear stellar disc
\citep[see][]{Launhardt:2002nx}. Fortunately, there are several dark
clouds contained in the field-of-view (FOV). Those clouds belong most
probably to dense gas and dust in the so-called circumnuclear ring
(CNR) in front of the nuclear star cluster \citep[see,
e.g.][]{Ekers:1983uq,Lo:1983kx,Christopher:2005fk} and can thus serve
to estimate the flux offset. We measured the median flux density at
six positions within these dark clouds (see blue circles in
Fig.\,\ref{Fig:wide}) and thus obtained an estimate for the mean and
standard deviation of the constant diffuse flux offset in the $K_{s}$
wide field image: $0.3\pm0.1$\,mJy\,arcsec$^{-2}$ (corresponding to
roughly $5$\,mJy\,arcsec$^{-2}$ if corrected for 3\,mag of extinction). The
surface flux density measurement was then repeated after subtracting
this mean offset. The extinction correction was performed after the
subtraction of this potential background bias. The resulting SB
profile and best fit model is shown in panel a) of
Fig.\,\ref{Fig:app4}, with the best-fit parameters listed in row 3 of
Tab.\,\ref{Tab:sys}. It can be seen that uncertainty about an additive
offset from the sky or from diffuse foreground radiation can have a
significant (order $20\%$) effect on the measured value of $\Gamma$ --
and, by consequence, also on $\beta$.

Since in this work we analyse observations with different filters and
instrument setups and taken under different conditions, we expect that
we can accurately estimate the contribution from the variability of
the atmospheric emission from the standard deviation of the results
for the different filters (at least for $\Gamma$). As concerns the possible
contribution of a diffuse component from Galactic structures in the
foreground of the NSC, in particular the nuclear disc, its effect will
always be a positive offset. That means that, if we subtract such an
offset, $\Gamma$ would increase.

\subsection{Extinction correction}

The strong differential extinction in the central parsecs of the Milky
Way is well known and we correct for it in our measurements. If we
assume simply a constant extinction and do not correct for its
variation, then the reduced $\chi^2$ becomes higher and the
gas-subtracted SFP profile can be  fit less well with a power-law. The
normalisation of the SB changes by $<10\%$ and $\Gamma$ becomes
steeper (see panel b) in Fig.\,\ref{Fig:app5} and row 2 in
Tab.\,\ref{Tab:sys}). The latter is to be expected because extinction
is lower near Sgr\,A* \citep[see, e.g. extinction maps presented
in][]{Schodel:2007tw,Schodel:2010fk}.

In any case, this is an extreme test that overestimates the
uncertainties probably significantly because, after all, interstellar
extinction and its variation towards the GC have been investigated well
and can be robustly estimated
\citep[e.g.][]{Schodel:2007tw,Buchholz:2009fk,Schodel:2010fk,Nishiyama:2013fk,2015ApJ...813...27H,Fritz:2016fj}.
We therefore repeated our analysis twice, once with the extinction map
smoothed by a Gaussian of $2"$ FWHM and once with the extinction map
smoothed by a median filter in a box of  of $4"$ width. In these cases
the final results agree, within the uncertainties, with our best
estimate. We conclude that the correction for variable interstellar
extinction is not a significant source of systematic error in this work.

\subsection{Completeness effects}

When studying stellar number densities, as in Paper I, assessing and
correcting incompleteness due to sensitivity and, in particular,
crowding can have significant effects on the results. In our study of the
diffuse light density, bias related to completeness could occur as
well. The contribution of the occasional bright star on the mean
surface brightness at a given $R$ will be negligible due to our use of
the ROBUST\_MEAN procedure, that rejects outliers and produces values
very similar to the median. However, in small, crowded areas, such as the
central arcseconds near Sgr\,A*, subtraction of faint stars may be
significantly less complete so that, on average brighter stars remain
in the image than in less crowded areas, which may create a systematic effect.

To examine this effect, we studied the SFP profile in images, in which
stars down to different magnitude levels were subtracted: $K_{s}=16$,
$K_{s}=18$, and all detectable stars. The $3\,\sigma$ detection limit
for stars in the $K_{S}$ wide field image is about $K_{S}\approx19$
(albeit at low completeness). The resulting profiles and best fits are
shown in panel c) of Fig.\,\ref{Fig:app5}. The corresponding best-fit
parameters are listed in rows 5 and 6 of Tab.\,\ref{Tab:sys}.  Apart
from an overall $\sim$$10-20\%$
shift between the measured SBs, the profiles look very similar. The
best fit parameters - apart from the SB normalisation,
$\Sigma_{0}$,
- show only a small range of bias. In particular, no significant
change of the best-fit parameters occurs whether we subtract all
detectable stars or only stars down to $K_{s}=18$.
We conclude that completeness effects are not any significant source of
systematic error in this work.

\subsection{Masking}
As shown in Fig.\,\ref{Fig:wide}, we mask several regions, that is, we
exclude them from the analysis. These regions are extended dark
clouds, residuals near the brightest star (IRS\,7), objects with
strong excess from line-emission or hot dust (e.g. IRS\,1W, IRS\,21,
or IRS\,13), or negative residuals around the densely clustered bright
stars near Sgr\,A*. As panel d) in Fig.\,\ref{Fig:app5} and row\,4 in
Tab.\,\ref{Tab:sys} show, suppression of masking makes the fit
noisier, but does not alter the best-fit parameters significantly. We
conclude that the choice of masking applied in this work is not any
significant source of systematic error.

\subsection{Binning}

\begin{figure*}[!htb]
\centering
\includegraphics[width=.8\textwidth]{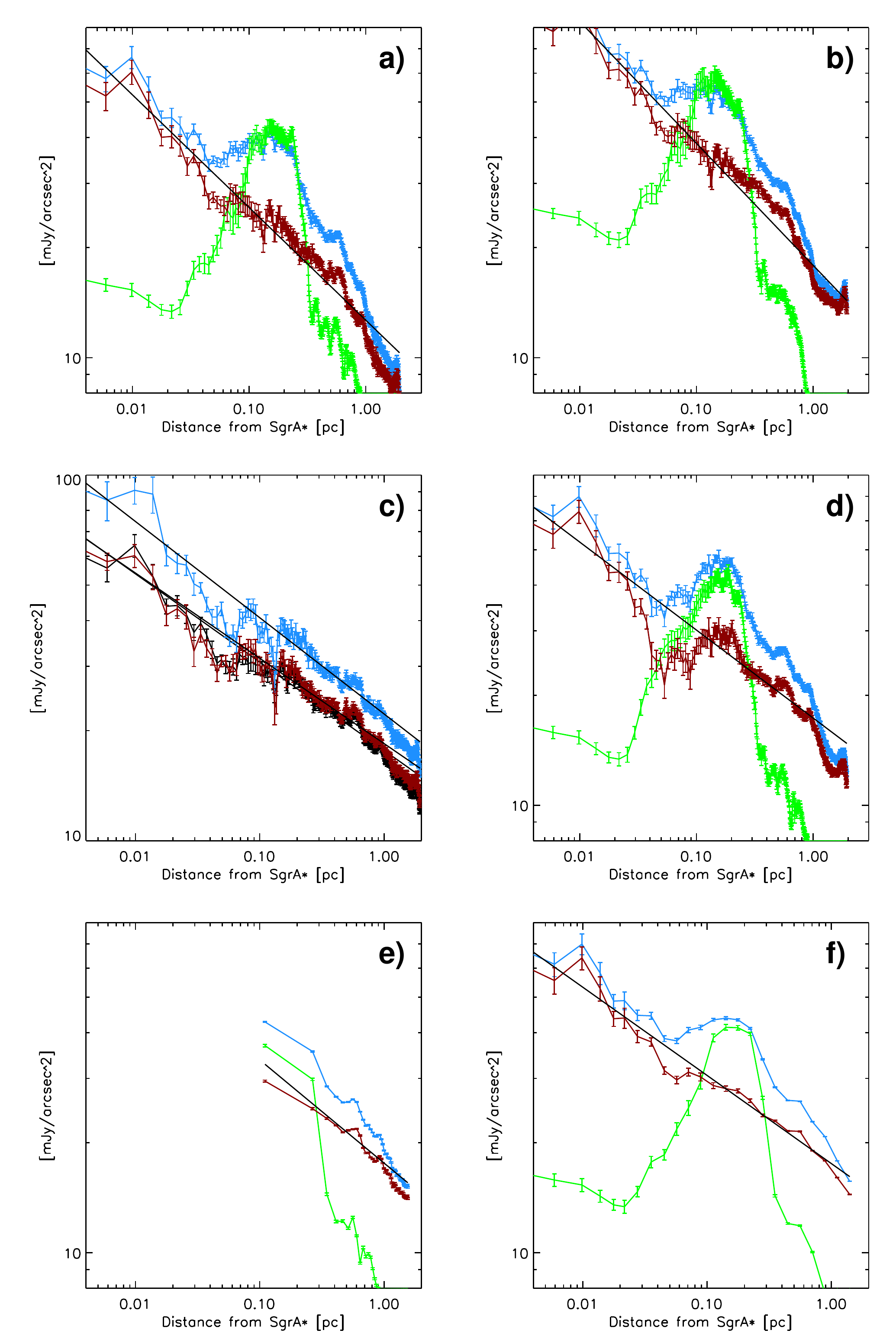}
\caption{\label{Fig:app5} Fits of the SB profile from the $K_{s}$
  wide field image to test potential sources of systematic errors. a)
  Subtraction of a potential sky offset. b) Assumption of constant
  extinction across the field. c) Subtraction of point sources down to
  $K_{s}=16$ (blue), $K_{s}=18$ (red), and all detectable point
  sources (back). d) No masking of dark clouds, or of systematic
  positive or negative residuals. e) Binning with a constant number of
  pixels ($1\times10^{4}$) per bin. f) Logarithmic binning, leading to
  a higher weight of the inner bins. }
\end{figure*}

We examined two different ways of binning the data. First, we binned
the data in a way that each bin contained the same number of
pixels. This will mean that the bins become smaller at larger $R$. A
small ($\sim$15$\%$) increase of $\Gamma$ is observed (see panel e) in
Fig.\,\ref{Fig:app5} and row\,7 in Tab.\,\ref{Tab:sys}). However, we
have chosen an extreme case of binning ($1\times10^{4}$ pixels or
about 100\,arcsec$^{2}$ per bin), which eliminates all data pints at
$R<0.1$\,pc. For a less extreme binning of 10\,arcsec$^{2}$ per bin,
the differences in the best-fit parameters are much smaller, with
$\Gamma=0.25$.

We also tested logarithmic binning, which results in increasingly
larger bins for larger $R$. As shown in panel f) in
Fig.\,\ref{Fig:app5} and row\,8 in Tab.\,\ref{Tab:sys}, this produces
no significant deviation in the best-fit parameters.

We conclude that binning is probably not any significant source of
systematic errors in this analysis, but may contribute an uncertainty
on the order of $\sim$$5\%$ to the value of $\Gamma$.

\begin{table*}
\centering
\caption{Best-fit parameters for $Ks$ wide-field image, under
  different circumstances that may affect systematics. We note that all
  the formal uncertainties of the best-fit parameters are
  $\leq1\%$, with the exception of row 8, which has significantly
  larger formal uncertainties due to the large reduced $\chi^{2}$. The
  $\chi^{2}$  listed here are smaller than the ones listed in
  Tab.\,\ref{Tab:Gammas} because the fitting range is different here
  ($R\leq0.5$\,pc compared to $R\leq 1$\,pc in the main body of the
  paper; see discussion on the change of the projected power-law in
  sections\,\ref{sec:range} and \ref{sec:nuker}). Therefore they are not listed in this table, which serves
  to explore systematic uncertainties, which dominate the error budget.}
\label{Tab:sys}
\begin{tabular}{lllll}
\hline
\hline
& $\Sigma_{0}$ & $\Gamma$ & $\Sigma_{0}$ & $\chi^{2}_{red}$\\
& (mJy\,arcsec$^{-2}$) & & & \\
\hline
1 & $14.3$ & $0.32$ & $0.057$& $0.3$ \\
2 & $22.6$ & $0.33$ & $0.045$ & $1.0$ \\
3 & $15.7$ & $0.31$ & $0.057$& $0.5$ \\
4 & $20.5$ & $0.24$ & $0.062$ & $0.3$ \\
5 & $26.8$ & $0.25$ & $0.058$ & $0.1$ \\
6 & $21.7$ & $0.23$ & $0.057$ & $0.4$ \\
7 & $20.7$ & $0.27$ & $0.062$& $0.5$ \\
8 & $21.4$ & $0.28$ & $0.054$& $2.5$ \\
 \hline
\end{tabular}
\begin{list}{}{}
\item[1] Final product as used in the results of this paper: Masking and extinction correction applied.
\item[2] No correction of differential extinction, assumption of $A_{Ks} =3.0$ constant.
\item[3] Masking and extinction applied. Subtraction of potential sky
  offset of $0.3$\,mJy\,arcsec$^{-2}$.
\item[4] No masking applied.
\item[5] Stars only subtracted if they are brighter than $K_{s}=16$.
\item[6] Stars only subtracted if they are brighter than $K_{s}=18$.
\item[7] Like 1, but using bins with constant number of pixels ($1\times10^{4}$) per bin.
\item[8] Like 1, but using bins of equal logarithmic width.
\end{list}
 \end{table*}

\subsection{Fitting range}
\label{sec:range}
\begin{figure}[!htb]
\includegraphics[width=\columnwidth]{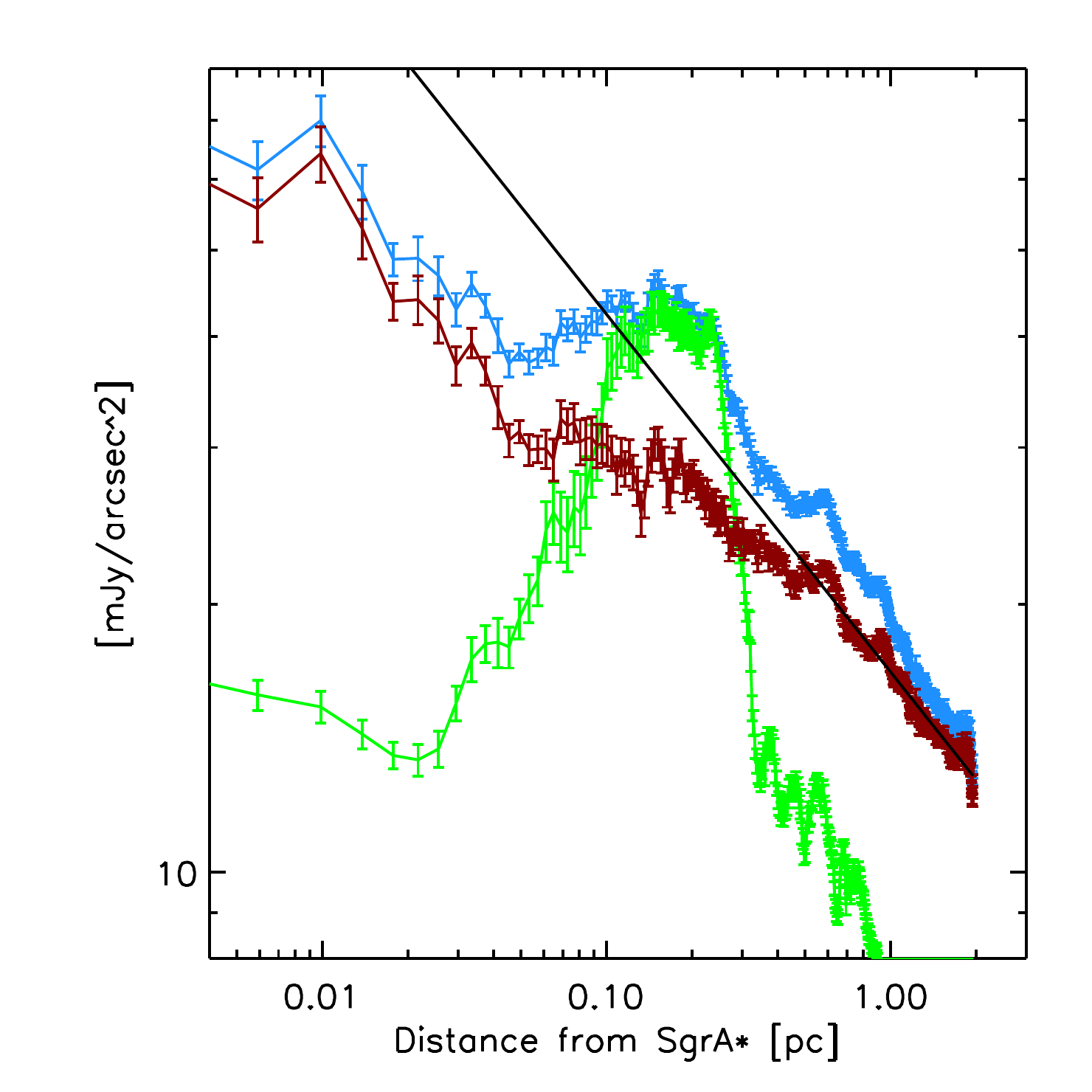}
\caption{\label{fig:app6} Mean diffuse SB profiles in the
  $K_{S}$ wide-field image before (blue) and after (red) subtraction of
the scaled Pa\,$\alpha$ emission (green). the straight black line is a
simple power-law fit to the data at $R\geq0.5$\,pc.}
\end{figure}

Finally, we study the role of the range in $R$ used to fit the
power-law from the stellar diffuse emission. When we only include data
at $R\leq0.4,0.6,0.8,1.0,1.2,1.5$\,pc, we obtain
$\Gamma=0.22,0.21,0.23,0.24,0.27,0.31$, and
$\Sigma_{0}=21.3,21.7,21.3,20.9,20.4,20.3$\,mJy\,arcsec$^{-2}$. As we
can see, there is a systematic effect with the power-law becoming
steeper at larger $R$. If we fit only data at
$R\geq0.5$\,pc, then we obtain
$\Gamma=0.40$ and $\Sigma_{0}=22.2$\,mJy\,arcsec$^{-2}$. We show the
corresponding fit in Fig.\,\ref{fig:app6}. On the other hand, if we
fix the outer edge of the fitting range to $R=1.0$\,pc and then use only
data at $R\geq0.2,0.4,0.6$\,pc, we obtain $\Gamma=0.30,0.33,0.35$, and
$\Sigma_{0}=21.5,21.7,22.0$\,mJy\,arcsec$^{-2}$.

We conclude that the data from the $K_{s}$ wide field image show
evidence for a steepening of the power-law with increasing $R$. When
we analyse the SB profiles for the different observations in this
work, we will always fit the power-law in the range
0\,pc$\leq R\leq$1\,pc. From our analysis here we estimate that the
corresponding best-fit values of $\Gamma$ may have an associated
systematic uncertainty on the order of $0.05$. The fitting range has
only a minor contribution to the uncertainty of $\Sigma_{0}$, on the
order of $3\%$.

\end{appendix}

\subsection{Conclusion on systematic uncertainties}

From the study of the different potential sources of systematic
errors in this section we identify three effects with possibly
significant contribution: 1) An unknown additive sky offset, 2) the fitting
range, and 3) binning.  Effect 1) will, however, be absorbed by our
using of several independent data sets. It will mainly be important in
the sense of any contribution of a non-nuclear stellar population to
the diffuse light and then always act to increase the estimated
$\Gamma$. Effect 2) may contribute with a systematic error of $0.05$,
compared to at most $0.02$ from 3),
and will therefore dominate the budget of systematic errors. We will
adopt $0.05$ as our systematic uncertainty for $\Gamma$.

As concerns the normalisation of the diffuse flux density,
$\Sigma_{0}$ we cannot compensate potential atmospheric effects
through the use of different filters. We therefore consider that a $25\%$
systematic uncertainty may be a good estimate of the systematic
uncertainty for this parameter (see above). The effects of
binning and of fitting range can be neglected for this parameter.

\end{document}